\documentclass[showpacs,showkeys,superscriptaddress,aps]{revtex4}
\def\be{\begin{eqnarray*}}
\def\ee{\end{eqnarray*}}
\def\pr2{\prod_{r=\frac{1}{2}}^{\infty}}
\def\su2{\sum_{r=\frac{1}{2}}^{\infty}}
\begin{document}
\begin{flushright}
TIT/HEP-525 \\
{\tt hep-th/0408149} \\
August,\hspace{1mm}2004 \\
\end{flushright}
\title{D-brane correlators as solutions of Hirota-Miwa equation}
\author{Ryuichi SATO}
\email[email : ]{ryu@th.phys.titech.ac.jp}
\affiliation{Department of Physics, Tokyo Institute of Technology,\\ Tokyo, 152-8551 Japan}
\author{Satoru SAITO}
\email[email : ]{saito_ru@nifty.com}
\affiliation{Hakusan 4-19-10, Midoriku, Yokohama 226-0006 Japan}
\begin{abstract}
We present a tachyon field, which simply connects to the calculus of the tachyon condensation. The tachyon field acts on any bare boundary state: the Neumann or the Dirichlet state, and generates in a special case the boundary state suggested by S.P. de Alwis, which leads to the correct ratio between D-brane tensions. On the other hand we generalize the Hirota-Miwa equation to the case where there are two boundaries. We show that correlation functions made from only the integrand of the tachyon field satisfy the generalized Hirota-Miwa equation. Using the formulation based on this evidence, we suggest that the coordinates in which there exist tachyons on an unstable D-brane be identified with the soliton coordinates in integrable systems. We also evaluate these correlation functions, which have not yet been integrated, to obtain the local information about the tachyons on unstable D-branes. We further see how these amplitudes are affected through the tachyon condensation by integrating the correlators over the tachyon momenta and taking the on-shell limit.
\end{abstract}
\maketitle
\section{Introduction}
Recent studies of string theory have been focused on understanding of the dynamics of D-branes \cite{Pol} through the tachyon condensation \cite{Sen} and explored that tachyon is an essential object in the nonperturbative description of superstring theory. In conventional Type II string theories the tachyon modes, which cause instability of the theory, were GSO-projected out from the outset. It was shown, however, that tachyon itself did not imply inconsistency of the theory and the supersymmetric theory with some open tachyonic modes could be defined consistently. Rather it is more convenient for us to adopt this setup as a starting point because the existence of open tachyonic modes naturally introduces unstable D-branes in the theory \cite{Sen, Witten, Horava}. In this way, the tachyon has been extensively recognized as an essential object in the string theory and thus we have to understand the effect of the tachyons on unstable D-branes and plainly reproduce the phenomenon. 

For this purpose, the following boundary state has been suggested by S.P. de Alwis \cite{SPA}:
\begin{equation}
|B,u\rangle_{NSNS}=\prod_{n=1}^{\infty}\prod_{r=\frac{1}{2}}^{\infty}\frac{(1+\frac{u}{r})}{(1+\frac{u}{n})}\exp\!\bigg[\frac{1}{n}\frac{u-n}{u+n}\alpha_{-n}\widetilde{\alpha}_{-n}+i\frac{u-r}{u+r}\psi_{-r}\widetilde{\psi}_{-r}\bigg]|0\rangle \int_{-\pi R}^{\pi R}\exp\!\bigg[-\frac{u}{4}(x_{0})^{2}\bigg]|x_{0}\rangle\hspace{0.5mm}.
\label{boundary state by de Alwis}
\end{equation}
It is found that this boundary state becomes the Neumann state at $u=0$ and the Dirichlet one at $u=\infty$. Furthermore the overlap with the one graviton state $|g\rangle:=\psi_{-\frac{1}{2}}\tilde{\psi}_{-\frac{1}{2}}|0\rangle$
\begin{equation}
\prod_{\mu\in\parallel}\prod_{\mu\in\bot}\langle g|B,u\rangle_{NSNS}=C\prod_{\mu\in\parallel}(2\pi R^{\mu})\prod_{\mu\in\bot}F(u_{\mu})\int_{-\pi R}^{\pi R}\exp\!\bigg[-\frac{u_{\mu}}{4}(x_{0}^{\mu})^{2}\bigg]
\end{equation}
simply leads to the correct ratio between D-brane tensions
\begin{equation}
\frac{T_{p}}{T_{p-1}}=\frac{1}{2\pi\sqrt{\alpha^{\prime}}}\hspace{0.5mm}.
\label{D-brane tensions ratio}
\end{equation}

In spite of many successful results on tachyon condensation, however, there have been not known a convincing theory which is to describe dynamics of D-branes. The matrix models, for instance, have been proposed some years ago as candidates of such theory. But they are still far from our satisfaction and the M-theory itself remains mysterious.

The main purpose of this paper is to point out a strong correlation of the D-brane dynamics with the theory of integrable systems. Especially we derive a bilinear difference equation of Hirota type, which must be satisfied by every correlation function of tachyon fields. This is a generalization of the result known for the correspondence between open string correlation functions and the KP-hierarchy \cite{S}. 
It will be worthwhile to emphasize here that the integrable system we concern in this paper does not refer to any particular soliton equation but the totality of such equations. For example, the single Hirota's bilinear difference equation (HBDE), given by \cite{Hirota, Miwa} 
\begin{eqnarray}
& &\alpha\hspace{1mm}\tau(k_{1}-1,k_{2},k_{3})\hspace{1mm}\tau(k_{1},k_{2}-1,k_{3}-1) \nonumber \\
&+&\beta\hspace{1mm}\tau(k_{1},k_{2}-1,k_{3})\hspace{1mm}\tau(k_{1}-1,k_{2},k_{3}-1)  \label{original HBDE} \\
&+&\gamma\hspace{1mm}\tau(k_{1},k_{2},k_{3}-1)\hspace{1mm}\tau(k_{1}-1,k_{2}-1,k_{3})=0\hspace{1mm} \nonumber
\end{eqnarray}
with $\alpha+\beta+\gamma=0$, includes infinitely many soliton equations in the KP-hierarchy corresponding to various possible continuous limits of independent variable $k$'s. 
This equation is now widely called the \textit{Hirota-Miwa equation} and plays an essential role not only in the theory of solitons but also in statistical physics and quantum field theory \cite{KLWZ}. Since the function $\tau$ can be given explicitly, this nonlinear system is completely integrable \cite{MSato, DJKM}.

Before starting with the discussion in detail let us briefly explain what we mean by the correspondence between the D-brane dynamics and integrable systems. 
There have been some papers \cite{Kazakov}, which studied the Matrix model in connection with equation (4). Our approach, however, will be different from them. 
It is essentially important to incorporate fermionic sector into our argument to obtain correct results. We do not include its contribution in this section, however, to avoid kinematical complications and explore the essence of our argument. 

First we introduce a tachyon field
\begin{equation}
\Phi(X)=\int{\mathcal D}K \hspace{1mm}\tilde{\Phi}(K) V(K)\hspace{1mm},
\label{tachyon field1}
\end{equation}
where ${\mathcal D}K$ is a functional integral over momentum distribution $K$ of strings and $\tilde{\Phi}(K)$ is a weight function. $V(K)$ is the tachyon vertex operator associated with $K$. It becomes the familiar single tachyon vertex $:\exp[ik_jX(\sigma_j)]:$ with $X(\sigma_j)$ being the string coordinate of a particle of momentum $k_j$ as we specify $K$ by
\begin{equation}
K(\sigma)=2\pi k_j\delta(\sigma-\sigma_j).
\label{K}
\end{equation}
If we choose the weight function $\tilde\Phi(K)$ properly and denote the Dirichlet ($\rho=+1$) and the Neumann ($\rho=-1$) boundary states by $|\rho\rangle$, we can simply reproduce the boundary state (\ref{boundary state by de Alwis}) of the S.P. de Alwis type as (see (\ref{tachyon states for bosonic part}))
$$ \Phi(X)|\rho\rangle. $$

We consider the correlation function of tachyon vertices defined by
$$ \langle\rho|V(K_1)V(K_2)\cdots V(K_N)|\rho'\rangle. $$
When there is no boundary, {\it i.e.}, the case of $\rho=\rho'=0$, this function represents strings propagating free space and satisfies \cite{S} the Hirota-Miwa equation (\ref{original HBDE}). If there are boundaries the equation (\ref{original HBDE}) is modified by an additional boundary term as we will discuss later. 

The most important observation in this paper is the identification of D-brane coordinates with the soliton coordinates in integrable systems. 
Specifically we will show in \S III that the expectation values of the space-time coordinates $X$ on the D-brane along the Dirichlet directions and the `dual coordinates' along the Neumann directions are identified with the soliton coordinate $\xi(z)$ whose Fourier components are the soliton variables $t_n$'s , {\it i.e.},
\begin{equation}
\xi(z)=-\sum_{n=1}^\infty t_{n} z^n. 
\label{soliton coordinate}
\end{equation}
Throughout this paper we employ the complex variables $z$ and $\bar z$ when we discuss integrable systems while $\sigma$ and $\tau$, which are related by $z=e^{\tau-i\sigma}$, are used in the discussion of string theory. 
In the Neumann case the expectation value of $X$ is $x_0$, the center of mass coordinate, but the one of T-dual coordinate turns out to be $\xi(z)$. The soliton variables $t_n$'s in $(\ref{soliton coordinate})$ and $k_{j}$'s in (\ref{original HBDE}) are related to each other via the well-known Miwa transformation \cite{S, Miwa}
\begin{equation}
t_0=\sum_{j=1}^M k_j\ln z_j,\qquad t_{n}=\frac{1}{n}\sum_{j=1}^{M}k_{j}z_j^{-n}\quad (n=1,2,...)
\label{Miwa transformations}
\end{equation}
when the momentum distribution $K(\sigma)$ is given by a sum of (\ref{K}) over $j$. Since we need not specify the number $M$ in the summation of (\ref{Miwa transformations}) we will not write it explicitly in what follows.

In terms of $\xi$ one soliton solution, for example, is simply given by $\tau=1+e^\xi$, while the periodic solutions with arbitrary number of genus are given by \cite{S, KS}
$$
\tau=\exp\left[-\ {1\over 8\pi^2}\oint d\xi(z)\oint d\xi(z'){E(z, z')\over z-z'}\right]{\theta\left(\phi_0+\oint \xi(z)\omega\right)\over\theta(\phi_0)},
$$
where $\theta(\phi),\ E(z,z')$ and $\omega$ are the Riemann theta function, prime form and the 1st Abel differential associated with a Riemann surface, respectively.

Based on this observation we see that the generalized Hirota-Miwa equation with the boundary term governs the reaction of the correlation function under the change of the path on a D-brane along which the strings are attached. 
At the same time this new interpretation of the momentum distribution enables us to understand the meaning of the Miwa transformation, which remained unclear in our previous works \cite{SS}. \\[1mm]

This paper is organized as follows. 
In \S II we introduce a tachyon field (\ref{tachyon field1}), which simply connects to the calculus of the tachyon condensation and show that this field reproduces the boundary state by S.P. de Alwis (\ref{boundary state by de Alwis}) from the bare (either the Neumann or the Dirichlet) boundary state. \S II-A is bosonic part and \S II-B is fermionic one (NSNS sector).
\S III is the main part of this paper in which we discuss the relations between D-branes and integrable systems. In \S III-A we derive a generalization of the original HBDE (\ref{original HBDE}) to the case where there are two boundaries. In \S III-B,C using the formulation of the tachyon field based on the previous evidence, we show that the coordinates of tachyons on an unstable D-brane can be identified with the soliton coordinates in integrable systems. 
In \S IV-A,B we calculate the correlation functions, which have not yet been integrated, to obtain the local information about the tachyons on unstable D-branes. Moreover, in \S IV-C, by integrating the correlators over the tachyon momenta and taking the on-shell limit, we see how these amplitudes are affected through the tachyon condensation. 
Correspondingly we classify the decay of unstable D-branes, which is known as the \textit{descent relations} \cite{Sen}. All calculus in this section is the one-dimensional case but prepare for the next section.
In \S V we consider the physics in the ten-dimensional space-time by collecting the results in the previous sections, and close this paper with summary and remarks in \S VI. 
In Appendix A we present formulae to help calculations in \S III, IV.
\section{Tachyon Field}
In this section we introduce a tachyon state which simply connects to the calculus of the tachyon condensation. We begin with the bosonic part in \S II-A and then directly extend to the fermionic part (NSNS sector) in \S II-B. At the end of this section it will be shown explicitly that our tacyon state reproduces the boundary state by S.P.de Alwis \cite{SPA} as a special case.
\subsection{Bosonic Part}
We first present a natural generalization of the single tachyon vertex operator to the one which has arbitrary values of momenta and number of tachyons
\begin{equation}
:e^{ik\cdot X}: \hspace{5mm} \longrightarrow \hspace{4mm} V(K)\hspace{0.5mm}=\hspace{1mm}:e^{i(K,X)}:\hspace{0.5mm},
\label{tachyon vertex op.}
\end{equation}
where $:\ :$ means the normal ordering and we define
\begin{equation}
i(K,X)\equiv i\int_0^{2\pi}{d\sigma\over 2\pi}\int_0^{2\pi}{d\sigma'\over 2\pi}K(\sigma)\hspace{0.5mm}\ln\left|e^{i\sigma}-e^{i\sigma'}\right|\hspace{0.5mm}{\partial X(\sigma')\over\partial\sigma'}.
\label{i(K,X)}
\end{equation}
We notice that the vertex operator of this form was first introduced to describe interaction of three strings \cite{DSS}. It is now going to be used again to describe an interaction of strings attached on a D-brane.

For simplicity we consider only one coordinate direction and omit space-time indices until \S V. Since we are interested in the nature of duality we introduce an index $\zeta$ and distinguish the closed string coordinate whether it is in the ordinary theory ($\zeta=1$) or in the T-dual theory ($\zeta=-1$). Hence the space-time coordinate of closed string is written as 
$$
X^\zeta(\sigma,\tau)=X(e^{\tau-i\sigma})+\zeta\tilde X(e^{\tau+i\sigma}).
$$
describing the ordinary theory with radii $R$ if $\zeta\!=\!+1$ and the one after taking the T-duality with radii $R^{\prime}=\alpha^{\prime}/R$ if $\zeta\!=\!-1$. The string coordinate on the boundary at $\tau=0$ is expanded according to
$$
X^\zeta(\sigma,0)=X_+^\zeta(\sigma,0)+X_-^\zeta(\sigma,0)
$$
\begin{equation}
X_\pm^\zeta(\sigma):=X_\pm^\zeta(\sigma,0)
=
\left(\matrix{x_0\cr 0\cr}\right)\mp i\sum_{n=1}^\infty{1\over n}\left(\alpha_{\mp n}e^{\mp in\sigma}+\zeta\tilde\alpha_{\mp n}e^{\pm in\sigma}\right).
\label{mode expansion for bosonic part}
\end{equation}

Correspondingly we also expand the distribution functions $K$ of tachyon momentum on a D-brane as
\begin{equation}
K(\sigma)=\sum_{n=1}^{\infty}(p_{n}e^{-in\sigma}+\bar{p}_{n}e^{in\sigma}).
\label{tachyon momentum}
\end{equation}
Using these components we define the weight function by
\begin{equation}
\tilde{\Phi}_{u}(K)=\prod_{n=1}^{\infty}\exp\!\bigg[-\frac{1}{2n(c_{n}^{u})^{2}}p_{n}\bar{p}_{n}\bigg]. 
\label{momentum rep. of tachyon field}
\end{equation}
We introduce a real parameter $u$ as a parameter which interpolates various types of boundary state.  
We do not need to specify the coefficient $c_{n}^{u}\equiv c_{n}(u,\rho)$ as a function of $u$ and $\rho\hspace{0.6mm}(=\pm 1)$ within our formulation but require it to reproduce the results of S.P. de Alwis (\ref{boundary state by de Alwis}) under certain condition. 

Let us denote by $\Phi_u(X^\zeta)$ the $u$-dependent tachyon field, {\it i.e.},
\begin{equation}
\Phi_{u}(X^\zeta)=\int{\mathcal D}K \hspace{1mm}\widetilde{\Phi}_{u}(K) :e^{i(K,X^\zeta)}:\hspace{1mm}.
\label{tachyon field2}
\end{equation}
This has an equivalent expression obtained by the change of momentum variables 
\begin{equation}
K \rightarrow K^{\prime}\hspace{5mm}(p_{n} = c_{n}^{u}p^{\prime}_{n}\hspace{1mm},\quad \bar{p}_{n} = c_{n}^{u}\bar{p}_{n}^{\prime}),
\end{equation}
as (we write $K$ in the place of $K'$)
\begin{equation}
\Phi_{u}(X^\zeta)=\prod_{n=1}^{\infty}(c_{n}^{u})^{2}\int\prod_{n=1}^{\infty}i\frac{dp_{n} d\bar{p}_{n}}{4n\pi}\hspace{0.5mm}\exp\!\bigg[-\frac{1}{2n}p_{n}\bar{p}_{n}\bigg]\hspace{1mm}:e^{i(K,X^\zeta)_{u}}:\hspace{1mm}.
\label{tachyon state as our starting point}
\end{equation}
In this form the weight function is the standard Gaussian one. The $u$-dependence is transferred into $(K,X^\zeta)_{u}$, which is now given by 
\begin{equation}
i(K,X^\zeta)_{u}:=i\int_{0}^{2\pi}\frac{d\sigma}{2\pi}K(\sigma)\int_{0}^{2\pi}\frac{d\sigma^{\prime}}{2\pi}\frac{\partial X^\zeta(\sigma^{\prime})}{\partial\sigma^{\prime}}\Delta_{u}(\sigma,\sigma^{\prime}),
\label{phase1 of vertex op.}
\end{equation}
and
\begin{equation}
\Delta_{u}(\sigma,\sigma^{\prime}):=-\frac{1}{2}\sum_{n=1}^{\infty}\frac{c_{n}^{u}}{n}\left(e^{in(\sigma-\sigma^{\prime})}+e^{-in(\sigma-\sigma^{\prime})}\right). 
\label{deformed Green's function}
\end{equation}
is a deformed Green's function which turns back to $\ln\left|e^{i\sigma}-e^{i\sigma'}\right|$ of (\ref{i(K,X)}) at $c^u_{n}=1$.

Now we act the tachyon field (\ref{tachyon state as our starting point}) on the bare boundary state: either the Neumann or the Dirichlet state, which we represent as $|\rho\rangle$
\begin{equation}
|\rho\rangle:=\prod_{n=1}^{\infty}\exp\!\bigg[\frac{\rho}{n}\alpha_{-n}\widetilde{\alpha}_{-n}\bigg]|0\rangle \quad \cases{\rho=-1 \quad Neumann \cr \rho=+1 \quad Dirichlet \hspace{0.5mm}.\cr}
\label{bare boundary state}
\end{equation}
Due to the well known nature of the bare boundary states, which is simply given by
$$
\left(X^\zeta(\sigma,0)-x_0\right)|\rho\rangle=(1-\rho\zeta)X_-^\zeta(\sigma)|\rho\rangle
=
(1-\rho\zeta)\left(X_+^\zeta(\sigma)-x_0\right)|\rho\rangle
$$
in our terminology, we obtain
\begin{equation}
:e^{i(K,X^{\zeta})_{u}}:|\rho\rangle \\[2mm]
=\cases{\hspace{1mm}e^{2i(K,X_{+}^{\zeta})_{u}}\hspace{0.5mm}|\rho\rangle \hspace{1cm}(\zeta\rho=-1) \cr 
\hspace{1mm} |\rho\rangle \hspace{2.62cm} (\zeta\rho=+1). \cr}
\label{b-state}
\end{equation}

From this we can deduce an interesting feature of the boundary state $\Phi_u(X^\zeta)|\rho\rangle$. Namely the tachyons do not affect the Dirichlet (Neumann) boundary state if $\zeta=+1\ (-1)$. This property of our boundary state will play an important role in later discussions. Note that this holds irrespective of the detail of $c_n^u$. 

It is not difficult to reproduce the result of S.P. de Alwis by choosing $c_{n}^{u}$ as
\begin{equation}
c_{n}^{u}:=c_{n}(u,\rho)=\sqrt{\frac{1}{1+\left(\frac{u}{n}\right)^{\rho}}}=\cases{\sqrt{\frac{u}{u+n}}\quad (\rho=-1) \cr \sqrt{\frac{n}{u+n}}\quad (\rho=+1).}
\label{c_n^u}
\end{equation}
In this particular choice of $c_{n}^{u}$ the contribution from the bosonic part to the state (\ref{boundary state by de Alwis}) is
\begin{equation}
\Phi_{u}(X^{\zeta})|\rho\rangle=\prod_{n=1}^{\infty}\frac{1}{1+(\frac{u}{n})^{\rho}}\exp\!\bigg[\frac{1}{n}\frac{u-n}{u+n}\alpha_{-n}\widetilde{\alpha}_{-n}\bigg]|0 \rangle,
\label{tachyon states for bosonic part}
\end{equation}
when $\zeta\rho=-1$. This result will be discussed in the next subsection.
\subsection{Fermionic Part}
We consider here an extension of the previous subsection to the fermionic part. We may consider only NSNS sector since RR sector does not include tachyonic modes in the spectra. \\[2mm]

We again introduce an index $\zeta^\psi$ to specify the fermionic field $\psi(\sigma,\tau)$ whether it is in the ordinary theory $(\zeta^\psi=1)$ or in the T-dual theory $(\zeta^\psi=-1)$;
$$
\psi^{\zeta^\psi}(\sigma,\tau)=\psi(e^{\tau-i\sigma})+\zeta^\psi\tilde\psi(e^{\tau+i\sigma}).
$$
Accordingly the mode expansion of $\psi^{\zeta^\psi}(\sigma,0)=\psi^{\zeta^\psi}_+(\sigma)+\psi^{\zeta^\psi}_-(\sigma)$ at the boundary ($\tau=0$) is given by
\begin{equation}
\psi_{\pm}^{\zeta^\psi}(\sigma)=\sum_{r=\frac{1}{2}}^{\infty}(\psi_{\mp r}e^{\mp ir\sigma}+i\zeta^{\psi}\hspace{0.5mm}\widetilde{\psi}_{\mp r}e^{\pm ir\sigma}).
\end{equation} 

The extension of the arguments in \S II-A to the fermionic case is almost straightforward. We define the following.\\[2mm]
Boundary state for NSNS sector ($\eta$ the spin structure);
\begin{equation}
|\eta,\rho\rangle:=\prod_{r=\frac{1}{2}}^{\infty}\exp\!\bigg[i\eta\rho\hspace{0.7mm}\psi_{-r}\widetilde{\psi}_{-r}\bigg]|0\rangle_{NSNS}
\label{bare NSNS boundary state}
\end{equation}
Distribution of tachyon momentum on D-brane ;
\begin{equation}
\Theta(\sigma)=\sum_{r=\frac{1}{2}}^{\infty}(\theta_{r}e^{-ir\sigma}+\bar{\theta}_{r}e^{ir\sigma}) 
\end{equation}
Weight function of fermionic tachyon field ;
\begin{equation}
\widetilde{\Phi}_{u}(\Theta)=\prod_{r=\frac{1}{2}}^{\infty}\exp\!\bigg[-\frac{1}{2(c_{r}^{u})^{2}}\theta_{r}\bar{\theta}_{r}\bigg] 
\end{equation}
Deformed Green's function ;
\begin{equation}
\Delta_{u}^{NSNS}(\sigma,\sigma^{\prime})
=-\frac{1}{2}\sum_{r=\frac{1}{2}}^{\infty}c_{r}^{u}(e^{ir(\sigma-\sigma^{\prime})}+e^{-ir(\sigma-\sigma^{\prime})}) 
\end{equation}
Tachyon field ;
\begin{eqnarray}
\Phi_{u}(\psi^{\zeta^\psi})
&=&\prod_{r=\frac{1}{2}}^{\infty}\frac{1}{(c_{r}^{u})^{2}}\int{\mathcal D}\Theta\hspace{1mm}\widetilde{\Phi}(\Theta)\hspace{0.6mm}:\exp\!\bigg[i\hspace{0.6mm}(\Theta\hspace{0.6mm},\hspace{0.6mm}\psi^{\zeta^\psi})_{u}\bigg]: \nonumber \\[2mm]
&=&\int\prod_{r=\frac{1}{2}}^{\infty}\frac{2\hspace{0.7mm}d\theta_{r}d\bar{\theta}_{r}}{(c_{r}^{u})^{2}}\exp\!\bigg[-\frac{1}{2}\theta_{r}\bar{\theta}_{r}\bigg]\hspace{0.6mm}e^{i\hspace{0.6mm}(\Theta\hspace{0.6mm},\hspace{0.6mm}\psi_{+}^{\zeta^\psi})_{u}}\hspace{0.6mm}e^{i\hspace{0.6mm}(\Theta\hspace{0.6mm},\hspace{0.6mm}\psi_{-}^{\zeta^\psi})_{u}}
\label{tachyon field for fermionic part},
\end{eqnarray}
with
\begin{equation}
i(\Theta\hspace{0.6mm},\hspace{0.6mm}\psi_{\pm}^{\zeta^\psi})_{u}=\int_{0}^{2\pi}\frac{d\sigma}{2\pi}\int_{0}^{2\pi}\frac{d\sigma^{\prime}}{2\pi}\hspace{0.5mm}\Theta(\sigma)\hspace{0.5mm}\psi_{\pm}^{\zeta^\psi}(\sigma^{\prime})\hspace{0.5mm}\Delta_{u}^{NSNS}(\sigma,\sigma^{\prime}).
\end{equation}

Having prepared enough tools it is not difficult to see that $\Phi_{u}(\psi^{\zeta^\psi})$ of (\ref{tachyon field for fermionic part}) reproduces the fermionic part of the boundary state (\ref{boundary state by de Alwis}) by de Alwis in a special case. Due to the general property of $|\eta,\rho\rangle$
$$
\psi^{\zeta^\psi}(\sigma,0)|\eta,\rho\rangle=(1+\zeta^\psi\eta\rho)\psi^{\zeta^\psi}_+(\sigma)|\eta,\rho\rangle,
$$
the relation
\begin{equation}
:e^{i(\Theta,\psi^{\zeta^{\psi}})_{u}}:|\eta,\rho\rangle
=
\left\{\begin{array}{ll}
e^{2i(\Theta,\psi_+^{\zeta^{\psi}})_{u}}|\eta,\rho\rangle& \quad (\zeta^{\psi}\eta\rho=+1) \cr
|\eta,\rho\rangle&\quad (\zeta^{\psi}\eta\rho=-1) 
\label{fermionic relation}
\end{array}\right.
\end{equation}
holds corresponding to the bosonic relation (\ref{b-state}). When $\zeta^{\psi}\eta\rho=+1$ we substitute
\begin{equation}
c_{r}^{u}:=c_{r}(u,\rho)=\frac{1}{\sqrt{1+\left(\frac{u}{r}\right)^{\rho}}}=\cases{\sqrt{\frac{u}{u+r}} \quad (\rho=-1) \cr \sqrt{\frac{r}{u+r}} \quad (\rho=+1).}
\label{c_r^u}
\end{equation}
into $\Phi_{u}(\psi^{\zeta^\psi})$ and perform the functional integration over $\theta_r$'s and $\bar\theta_r$'s. We will then find the fermionic tachyon field $\Phi_{u}(\psi^{\zeta^\psi})$ also generates the fermionic part of the state (\ref{boundary state by de Alwis}) of S.P. de Alwis as
\begin{equation}
\Phi_{u}(\psi^{\zeta^{\psi}})|\eta,\rho \rangle
=\prod_{r=\frac{1}{2}}^{\infty}\left(1+\left(\frac{u}{r}\right)^{\rho}\hspace{0.5mm}\right)\exp\!\bigg[i\eta\frac{u-r}{u+r}\psi_{-r}\widetilde{\psi}_{-r}\bigg]|0\rangle_{NSNS}\hspace{0.5mm},
\label{tachyon states for fermionic part}
\end{equation}
as the counter part of the bosonic one (\ref{tachyon states for bosonic part}).
\subsection{Local and Global Tachyon States}

If we combine (\ref{tachyon states for bosonic part}) and (\ref{tachyon states for fermionic part}) we obtain
\begin{equation}
|B,\rho,\eta,u \rangle_{NSNS}=\prod_{n=1}^{\infty}\prod_{r=\frac{1}{2}}^{\infty}\frac{1+(\frac{u}{r})^{\rho}}{1+(\frac{u}{n})^{\rho}}\exp\!\bigg[\frac{1}{n}\frac{u-n}{u+n}\alpha_{-n}\widetilde{\alpha}_{-n}+i\eta\frac{u-r}{u+r}\psi_{-r}\widetilde{\psi}_{-r}\bigg]|0\rangle\otimes|0\rangle_{NSNS}\hspace{0.5mm}.
\label{tachyon state in SPA}
\end{equation}
In this formula we notice that the result of S.P. de Alwis (\ref{boundary state by de Alwis}) is fully realized when $\rho=+1$. In other words the theory of S.P. de Alwis corresponds to the theory with $\zeta=-1$ in our formulation. If we started our construction of the tachyon state from the Neumann boundary condition $(\rho=-1$), or the theory of $\zeta=+1$, we obtained the same state (\ref{boundary state by de Alwis}) again up to an overall constant factor.

From the result of (\ref{tachyon state in SPA}) we can calculate the vacuum expectation value of the full tachyon field $\Phi_{u}(X^{\zeta},\psi^{\zeta^\psi})$, which combines the bosonic (\ref{tachyon state as our starting point}) and the fermionic (\ref{tachyon field for fermionic part}) part. The full tachyon field is now given by 
\begin{equation}
\Phi_{u}(X^{\zeta},\psi^{\zeta^\psi})=\int{\mathcal D}K\hspace{0.5mm}{\mathcal D}\Theta\hspace{1mm}\widetilde{\Phi}(K)\hspace{0.5mm}\widetilde{\Phi}(\Theta):e^{i(K,X^{\zeta})_{u}+i(\Theta,\psi^{\zeta^\psi})_{u}}:\hspace{1mm}.
\label{full tachyon field}
\end{equation}
Its vev is easily evaluated, due to the conformal normal ordering and the normalization of the integral measures, as
$$
\langle\Phi_{u}\rangle:=\langle\hspace{0.5mm}\Phi_{u}(X^{\zeta},\psi^{\zeta^\psi})\hspace{0.5mm}\rangle=\prod_{n=1}^{\infty}\prod_{r=\frac{1}{2}}^{\infty}\frac{(c_{n}^{u})^{2}}{(c_{r}^{u})^{2}}.
$$
This corresponds to the maximum condensation of tachyons at $u=0$ $(u=\infty)$ according to the sign $\zeta=+1$ $(\zeta=-1)$, and explains why we call $\Phi_u$ the \textit{tachyon field}. 

Apart from these global informations discussed above we are mainly concerned with local informations of the tachyon fields of which we will make the most in the following sections. Namely we are interested in the states defined by
\begin{eqnarray}
|T_{u}^{\zeta}(K)\rangle&:=&e^{2i(K,X_{+}^{\zeta})_{u}}|\rho\rangle, \hspace{1.45cm}(\zeta\rho=-1)
\label{tachyon state |T>} \\[3mm] 
|T_{u,\eta}^{\zeta^{\psi}}(\Theta)\rangle&:=& e^{2i(\Theta,\psi_{+}^{\zeta^{\psi}})_{u}}|\eta,\rho\rangle, 
\hspace{10.3mm} (\zeta^{\psi}\eta\rho=+1) .
\label{tachyon states1}
\end{eqnarray}
We will call these states with the relation $\zeta\rho=-1$, $\zeta^{\psi}\eta\rho=+1$ \textit{tachyon states}. On the other hand an action of the tachyon vertex operator to the state $|\rho\rangle$, $|\eta,\rho\rangle$ has no effect when $\zeta\rho=+1$, $\zeta^{\psi}\eta\rho=-1$. We call such states \textit{non-tachyon states}.

The tachyon states (\ref{tachyon state |T>}) and (\ref{tachyon states1}) provide us local informations about the distribution of tachyon momenta at the boundary. As we will show in the next section the correlators of these states must satisfy a soliton equation. In other words it is the local state that is governed by the soliton equation.

\section{Correspondence with Integrable Systems}

This section constitutes the main part of this paper. Namely we derive explicitly a soliton equation satisfied by tachyon correlators between two D-branes. The equation, as it will be shown in the last part of \S III-A, is a difference equation which is a generalization of the Hirota-Miwa equation (\ref{original HBDE}) to the case where there are two boundaries. This result enables us to study the relation between D-branes and integrable systems. In \S III-B,C we show that the coordinates of tachyons on an unstable D-brane be identified with the soliton coordinates of integrable systems. 

\subsection{A Generalization of Hirota-Miwa Equation with two Boundaries}
The purpose of this subsection is to derive the equation (\ref{bilinear eq with boundary}), which determines behaviour of the $\tau$-functions. It has been known that the correlation functions of open bosonic strings satisfy the Hirota-Miwa equation (\ref{original HBDE}), a single discrete equation equivalent to the KP-hierarchy \cite{S}. In our previous paper we generalized it to the case of closed strings including a boundary \cite{SS}. Here we would like to show that we can further generalize the Hirota-Miwa equation to the case where there are two boundaries. This enables us to characterize correlation functions of closed strings propagating between two D-branes.

In order to avoid complication of kinematics we will again consider only the bosonic sector. Since the argument goes almost parallel to our previous work \cite{SS}, we do not repeat the details but begin with the identity which holds for an arbitrary element $G$ of the universal Grassmannian \cite{MSato, DJKM}
\begin{equation}
I:=\oint_0{dz\over 2\pi i}\oint_\infty{d\bar z\over 2\pi i}\phi^*(z,\bar z)G|0\rangle\otimes\phi(z,\bar z)G|0\rangle=0\hspace{1mm},
\label{I=0}
\end{equation}
where $\phi$'s are the fields related to $X(z,\bar z)$ through the bosonization
\begin{equation}
\phi(z,\bar z)=:e^{iX(z,\bar z)/\sqrt2}:\hspace{1mm},\quad \phi^*(z,\bar z)=:e^{-iX(z,\bar z)/\sqrt2}: \hspace{1mm}.
\label{closed string bosonization}
\end{equation}
The identity (\ref{I=0}) owes to the fermionic nature of these fields. We assume that the contours encircle around $z=0$ and $\bar z=\infty$. Since the difference of the sign of $\zeta$ plays no role in the derivation of soliton equation we will not write it throughout this subsection.

The propagation of solitons in the KP theory is generated by the Hamiltonian
$$ H(t):=-\sum_{n=1}^\infty\left(t_n\alpha_n+\bar t_n\widetilde\alpha_n\right). $$
The infinitely many variables $t_n$'s are the parameters along which solitons propagate according to the rules
\begin{eqnarray}
e^{H(t)}\hspace{0.7mm}\phi(z,\bar z)&=&e^{\xi(z,\bar z)/\sqrt 2}\hspace{0.7mm}\phi(z,\bar z)\hspace{0.7mm}e^{H(t)} \nonumber \\
e^{H(t)}\hspace{0.7mm}\phi^*(z,\bar z)&=&e^{-\xi(z,\bar z)/\sqrt 2}\hspace{0.7mm}\phi^*(z,\bar z)\hspace{0.7mm}e^{H(t)}
\label{e^Hpsie^H=e^xipsi}
\end{eqnarray}
where
\begin{equation}
\xi(z,\bar z)=-\sum_{n=0}^\infty\left(t_nz^n+\bar t_n\bar z^n\right)
\label{xi=tz}
\end{equation}
This $\xi(z,\bar z)$ is a generalization of the variable in (\ref{soliton coordinate}) which plays the central role in the theory of solitons, and we call it the {\it soliton coordinate}.

The correspondence of this theory of solitons with the theory of strings becomes clear through the Miwa transformaion (\ref{Miwa transformations}). In order to study the correspondence of integrable systems with closed strings we also introduce new variables $\bar t_n$'s by
\begin{equation}
\bar t_0=\sum_j k_j\ln\bar z_j,\qquad \bar t_n={1\over n}\sum_j k_j\bar z_j^{-n}\quad (n=1,2,...)
\label{Miwa' transformations}
\end{equation}
corresponding to the left mover.

In terms of new variables $k_j$'s the soliton coordinate $\xi$ and the Hamiltonian $H(t)$ are written as
$$ \xi(z,\bar z)=2\sum_jk_j\ln\left|z_j-z\right|,\qquad H(t)=i\sum_jk_jX_-(\sigma_j,\tau), $$
and the formula (\ref{e^Hpsie^H=e^xipsi}) is equivalent to
\begin{equation}
e^{H(t)}\hspace{0.7mm}\phi^*(z,\bar z)\otimes e^{H(t')}\hspace{0.7mm}\phi(z,\bar z)=\prod_j\left((z_j-z)(\bar z_j-\bar z)\right)^{(k'_j-k_j)/\sqrt 2}\phi^*(z,\bar z)\hspace{0.7mm}e^{H(t)}\ \otimes\ \phi(z,\bar z)\hspace{0.7mm}e^{H(t')}.
\label{e^Hpsie^H=e^xipsi'} 
\end{equation}
The variables $t'_n$ and $\bar t'_n$ in $H(t')$ we define simply replacing $k_j$ by $k'_j$ in (\ref{Miwa transformations}) and (\ref{Miwa' transformations}).

To apply the identity (\ref{I=0}) to our problem we consider $G$ being given by the boundary operator $T_u(K)$ in (\ref{tachyon state |T>}), {\it i.e.},
$$
T_u(K):=e^{2i(K,X_+)_u}\prod_{n=1}^\infty e^{(\rho/n)\alpha_{-n}\tilde\alpha_{-n}}.
$$
Since the exponent of this operator is either linear or bilinear in $\alpha_{-n}$ and $\tilde\alpha_{-n}$, it is guaranteed that $T_u(K)$ is an element of the Grassmannian. We multiply to $I$ the state
$$\left(\langle T(K')|:\exp\left[i\sum_jk_jX(z_j,\bar z_j)\right]:\right) \otimes \left(\langle T(K')|:\exp\left[i\sum_jk'_jX(z_j,\bar z_j)\right]:\right) $$
from the left. If we move the $X_-(z_j,\bar z_j)$ components to the right of $\phi$'s using (\ref{e^Hpsie^H=e^xipsi'}) we obtain
\begin{eqnarray}
0&=&
\oint_0{dz\over 2\pi i}\oint_\infty{d\bar z\over 2\pi i}\prod_{j}((z_j-z)(\bar z_j-\bar z))^{(k'_j-k_j)/\sqrt 2}\nonumber\\
&&
\langle T_v(K')|:\exp\left[i\sum_jk_jX(z_j,\bar z_j)-{i\over\sqrt 2}X(z,\bar z)\right]:|T_u(K)\rangle \nonumber\\
&\times&
\langle T_v(K')|:\exp\left[i\sum_jk'_jX(z_j,\bar z_j)+ {i\over\sqrt 2}X(z,\bar z)\right]:|T_u(K)\rangle \hspace{1mm}.
\label{0=...}
\end{eqnarray}

Since $k_j$'s and $k'_j$'s are arbitrary in (\ref{0=...}) let us choose them such that
$$ k'_j=k_j-\sqrt 2\quad (j=1,2,3) \qquad k'_j=k_j\quad (j\ne 1,2,3). $$
Then the product in (\ref{0=...}) turns simply to 
$$ \prod_{j=1,2,3}{1\over (z_j-z)(\bar z_j-\bar z)}. $$We assume that the two boundary states $|T_u(K)\rangle$ and $|T_v(K')\rangle$ are defined along two closed paths $\gamma, \gamma'$ on the complex $z$ plane which do not intersect each other but including the origin inside. We further assume that $z_1,z_2,z_3$ are inside of $\gamma$ and $\bar z_1,\bar z_2,\bar z_3$ are outside of $\gamma'$. Under this circumstance we move the contours of integration toward the boundaries. The contributions to the integration come from the simple poles at $z_j$'s and $\bar z_j$'s and those from the boundaries. 

The result can be summarized as follows:
\begin{eqnarray}
0&=&
{\tau^B(k_1-{1\over\sqrt2},k_2,k_3)\tau^B(k_1,k_2-{1\over\sqrt2},k_3-{1\over\sqrt2})\over|(z_1-z_2)(z_1-z_3)|^2}\nonumber\\
&+&
{\tau^B(k_1,k_2-{1\over\sqrt2},k_3)\tau^B(k_1-{1\over\sqrt2},k_2,k_3-{1\over\sqrt2})\over|(z_2-z_1)(z_2-z_3)|^2}\nonumber\\
&+&
{\tau^B(k_1,k_2,k_3-{1\over\sqrt2})\tau^B(k_1-{1\over\sqrt2},k_2-{1\over\sqrt2},k_3)\over|(z_3-z_1)(z_3-z_2)|^2}\nonumber\\
&+&
\oint_{\gamma}{dz\over 2\pi i}\oint_{\gamma'}{d\bar z\over 2\pi i}\prod_{j=1}^3{1\over (z-z_j)(\bar z-\bar z_j)}\nonumber\\
&&
\times \tau^B\left(k_1,k_2,k_3\right)\tau^B\left(k_1-{\scriptstyle{{1\over\sqrt2}}},k_2-{\scriptstyle{{1\over\sqrt2}}},k_3-{\scriptstyle{{1\over\sqrt2}}}\right).
\label{bilinear eq with boundary}
\end{eqnarray}
In this expression we introduced the $\tau$-function by
\begin{equation}
\tau^B(k_1,k_2,k_3):=\langle T_v(K')|:\exp\left[i\sum_jk_jX(z_j,\bar z_j)\right]:|T_u(K)\rangle \hspace{1mm}.
\label{tau}
\end{equation}
(\ref{bilinear eq with boundary}) is the generalization of the Hirota-Miwa equation to the case with two boundaries for which we have been looking.
It is important to notice that the two complex planes $z$ and $\bar z$ are connected dynamically by the boundaries $\gamma,\ \gamma'$ via reflections of strings.
\subsection{Role of Soliton Coordinate in String Theory --- bosonic part}
Based on our formulation in \S II and the evidence in \S III-A, we show in this subsection that the coordinates on a D-brane where tachyon fields are attached can be identified with the soliton coordinates $\xi(z,z^{-1})$ of (\ref{xi=tz}). To this end we first notice that the state (\ref{b-state}) with $\zeta\rho=-1$ is an eigenstate of either $X(\sigma,0)$ or $\partial X(\sigma,0)/\partial\tau$ depending on the sign of $\zeta$. In fact it satisfies
\begin{eqnarray}
X^{\zeta_{o}}(\sigma,0)|T_u^\zeta(K)\rangle &=& \xi_{c}^{X}(\sigma,0)|T_u^\zeta(K)\rangle,\hspace{1cm} (\rho\zeta_{o}=1) \label{X|>=xi|>}\\
{\partial X^{\zeta_{o}}(\sigma,0)\over\partial\tau}|T_u^\zeta(K)\rangle &=& {\partial \xi_{c}^{X}(\sigma,0)\over\partial\tau}|T_u^\zeta(K)\rangle,\hspace{7mm} (\rho\zeta_{o}=-1) \label{dX|>=dxi|>}
\end{eqnarray}
where
\begin{equation}
\xi_{c}^{X}(\sigma,\tau):=x_0-\sum_{n=1}^\infty{c_n\over n}\left(p_n e^{n(\tau-i\sigma)}+\bar{p}_{n} e^{-n(\tau-i\sigma)}\right),
\label{xi_c}
\end{equation}
and the lower index $c$ denotes coeffecients $\{c_{n}\}_{n\in\mathbf{Z}_{>0}}$ collectively, and the upper one $X$ means the bosonic part. 
In (\ref{X|>=xi|>}) and (\ref{dX|>=dxi|>}) the index $\zeta_{o}$ is written explicitly to specify the value $\pm 1$ in $X(\sigma,\tau)$ independent from the one in $|T_u^\zeta(K)\rangle$. It is clear from this result that the state with $\zeta\rho=+1$ is not an eigenstate of the coordinate $X(\zeta_{o}=\zeta)$ but an eigenstate of $X(\zeta_{o}=-\zeta)$. 
Along the Neumann direction $\rho=-1$, for example, the state is an eigenstate of the `dual coordinate' $X(\zeta_{o}=-1)$. Irrespective of the difference of the state, however, the eigenvalue $\xi_{c}^{X}$ itself is the same.

In terms of $\xi_{c}^{X}(\sigma):=\xi_{c}^{X}(\sigma,0)$ the function $i(K,X_+^\zeta)_u$ in the eigenstate $|T_u^\zeta(K)\rangle$ is then given by the expression
\begin{eqnarray*}
i(K,X_+^\zeta)_{u} 
&=& i\int_{0}^{2\pi}\frac{d\sigma}{2\pi}{\partial\xi_c(\sigma)\over\partial\sigma}\int_{0}^{2\pi}\frac{d\sigma^{\prime}}{2\pi}\frac{\partial X_+^\zeta(\sigma^{\prime},0)}{\partial\tau}\ln\left|e^{i\sigma}-e^{i\sigma'}\right| 
\label{equiv phase1 of vertex op.} \\
&=& {i\over 2\pi}\oint_\Gamma d\xi_c(\sigma)\int_{0}^{2\pi}\frac{d\sigma^{\prime}}{2\pi}\frac{\partial X_+^\zeta(\sigma^{\prime},0)}{\partial\tau}\ln\left|e^{i\sigma}-e^{i\sigma'}\right|,
\label{equiv phase1 of vertex op. 2}
\end{eqnarray*}
where $\Gamma$ denotes a path along which $\xi_{c}^{X}(\sigma)$ moves as $\sigma$ varies from 0 to $2\pi$. Since $\xi_{c}^{X}(\sigma)$ maps a point of the closed string on the boundary to a point in the space-time, $\Gamma$ is a path on the D-brane. This expression tells us that the eigenstate of the operator $X^{\zeta_{o}}(\sigma)$ is determined by specifying a path $\Gamma$ on the D-brane. We will call $\xi_{c}^{X}(\sigma)$ the {\it D-brane coordinate} (for bosonic part).

In order to see the correspondence of $\xi_{c}^{X}(\sigma)$ with the soliton coordinate $\xi(z,z^{-1})$ of (\ref{xi=tz}) we consider the case that the distribution of tachyon momentum $K(\sigma)$ is given by a sum of terms of (\ref{K}), {\it i.e.,}
$$ K(\sigma)=2\pi\sum_jk_j\delta(\sigma-\sigma_j). $$
It then follows that
$$ p_n=\sum_jk_je^{in\sigma_j},\qquad \bar{p}_{n}=\sum_jk_je^{-in\sigma_j},\qquad (n=1,2,...) $$
Substituting this result into (\ref{xi_c}) and comparing with (\ref{xi=tz}) we find
$$ \xi_{c}^{X}(\sigma)|_{c_n=1}=\xi(z,z^{-1}). $$
Hence the identification of the D-brane coordinate $\xi_{c}^{X}(\sigma)$ with the soliton coordinate has been established. In this correspondence, however, we must be careful. Namely the formulae (\ref{X|>=xi|>}) and (\ref{dX|>=dxi|>}) tell us that the string coordinate $X^{\zeta_{o}}(\sigma,0)$ with $\zeta_{o}=+1$ is diagonal only along the Dirichlet directions. In the Neumann directions the D-brane coordinate $\xi_{c}^{X}(\sigma)$ appears as an eigenvalue of the `dual coordinate' $X^{\zeta_{o}}$ with $\zeta_{o}=-1$. 

We can further confirm our assertion by the calculation of expectation values of $X^{\zeta_{o}}$. Note that we write the sign of the T-duality of $\zeta_{o}(=\pm 1)$ explicitly. 
Using {\tt [Formula\hspace{0.7mm}2]} in Appendix A,
$$ f_{n}:=\rho, \qquad g_{n}:=\rho e^{-2ns}, $$
$$ a_{n}=-i{c_{n}^{v}\over n}p_n,\quad \bar{a}_{n}=+i\zeta{c_{n}^{v}\over n}\bar{p}_{n}, \quad b_{n}=i{c_{n}^{u}\over n}\bar{p}_{n} e^{-ns},\quad \bar{b}_{n}=-i\zeta{c_{n}^{u}\over n}p_ne^{-ns},$$
we find
\begin{eqnarray}
\langle X^{\zeta_{o}}(\sigma)\rangle^\zeta
&\equiv&\lim_{s \to 0}\frac{\langle T_{v}^\zeta(K)|X^{\zeta_{o}}(\sigma)e^{-s(L_{0}+\widetilde{L}_{0})}|T_{u}^\zeta(K)\rangle}{\langle T_{v}^\zeta(K)|e^{-s(L_{0}+\widetilde{L}_{0})}|T_{u}^\zeta(K)\rangle}\bigg|_{u=v} \nonumber \\[2mm]
&=&\lim_{s \to 0}\prod_{n=1}^{\infty}\frac{{}_{f_{n}} \langle a_{n},\bar{a}_{n}|X^{\zeta_{o}}(\sigma)|b_{n},\bar{b}_{n}\rangle_{g_{n}}^\zeta}{{}_{f_{n}}\langle a_{n},\bar{a}_{n}|b_{n},\bar{b}_{n}\rangle_{g_{n}}^\zeta} \nonumber \\[2mm]
&=&x_0+{1-\zeta\zeta_{o}\over 2}\left(\xi_{c}^{X}(\sigma)-x_0\right) \nonumber \\[3mm]
&=&\cases{\xi_{c}^{X}(\sigma) \hspace{5mm}(\zeta\zeta_{o}=-1) \cr x_{0} \hspace{1.06cm}(\zeta\zeta_{o}=+1),} 
\label{<X>} \\[5mm]
\left\langle {\partial X^{\zeta_{o}}\over\partial\tau}\right\rangle^\zeta
&=&{1-\zeta\zeta_{o}\over 2}{\partial\xi_{c}^{X}(\sigma)\over\partial\tau}.
\label{<par X>}
\end{eqnarray}

Above statements are reproduced when $\rho\zeta_{o}\hspace{0.5mm}(\hspace{0.5mm}=-\zeta\zeta_{o})=+1$. This result also shows that the expectation value of the coordinate along the Neumann direction is simply the center of mass coordinate $x_0$, whereas that of momentum is given by $\partial\xi_{c}^{X}(\sigma)/\partial\tau$.

When the correspondence between string theory and integrable systems was first pointed out \cite{S}, it was already recognized that the variable $\xi$ of the soliton theory appeared in the theory of strings through the Miwa transformation. This correspondence had remained, however, rather mysterious since the idea of D-brane was not known and $\xi$ had no physical object in the string theory. The situation is now quite different. The formula (\ref{<X>}) clearly tells us that the eigenvalue $\xi_{c}^{X}(\sigma)$ of $X$ has the meaning of the place where the tachyon field is attached on the D-brane along the Dirichlet direction. Along the Neumann direction it is the expectation value of the `dual coodinate'. Irrespective of the direction in the space-time it plays the role of soliton coordinate in integrable systems.
\subsection{Role of Soliton Coordinate in String Theory --- fermionic part}
In this subsection, as well as \S III-B, we will calculate the following type of expectation values of $\psi$ on the tachyon states
\begin{equation}
\langle \psi(\sigma)\rangle^{\zeta^{\psi}}\equiv\lim_{s \to 0}\frac{\langle T_{v,\eta^{\prime}}^{\zeta^{\psi}}(\Theta)|\psi(\sigma)\hspace{0.5mm}e^{-s(L_{0}+\widetilde{L}_{0})}|T_{u,\eta}^{\zeta^{\psi}}(\Theta)\rangle}{\langle T_{v,\eta^{\prime}}^{\zeta^{\psi}}(\Theta)|e^{-s(L_{0}+\widetilde{L}_{0})}|T_{u,\eta}^{\zeta^{\psi}}(\Theta)\rangle}\bigg|_{u=v}.
\label{<psi>def}
\end{equation}
In particular we are interested in such a state which diagonalizes the field $\psi(\sigma)$. Since it is a Majorana field, the eigenvalues must be real. We will see that these requirements determine both the ket and bra states without ambiguity.

To this end we first act $\psi^{\zeta_{o}^{\psi}}(\sigma)$ to the fermionic tachyon state $|T^{\zeta^{\psi}}_{u,\eta}(\Theta)\rangle$ of (\ref{tachyon states1}). We introduced the sign $\zeta_{o}^{\psi}(=\pm 1)$ explicitly to emphasize that it could be different from $\zeta^{\psi}$ which specifies the state $|T^{\zeta^{\psi}}_{u,\eta}(\Theta)\rangle$. We then find, after some manipulation,
\begin{equation}
\psi^{\zeta_{o}^{\psi}}(\sigma)|T^{\zeta^{\psi}}_{u,\eta}(\Theta)\rangle=\xi_c^\psi(\sigma)|T^{\zeta^{\psi}}_{u,\eta}(\Theta)\rangle
\end{equation}
holds if and only if $\zeta^{\psi}\zeta_{o}^{\psi}=-1$. Here the fermionic eigenvalue function $\xi_c^\psi(\sigma)$ is given by
\begin{equation}
\xi_{c}^{\psi}(\sigma)=-\su2 c_{r}^u(\theta_{r}e^{-ir\sigma}+\bar{\theta}_{r}e^{ir\sigma}).
\end{equation}

The bra tachyon state $\langle T_{u,\eta}^{\zeta^{\psi}}(\Theta)|$ can be obtained from the ket state $|T^{\zeta^{\psi}}_{u,\eta}(\Theta)\rangle$ by conjugation:
\begin{eqnarray*}
\langle T_{u,\eta}^{\zeta^{\psi}}(\Theta)|
&:=&\langle \eta,\rho|:\exp\!\bigg[-i(\Theta,\psi^{\zeta^{\psi}})_{u}^{\dagger}\bigg]: \\[2mm]
&=&\langle \eta,\rho|e^{-i(\Theta,\psi_{-}^{\zeta^{\psi}})_{u}^{\dagger}}e^{-i(\Theta,\psi_{+}^{\zeta^{\psi}})_{u}^{\dagger}},
\end{eqnarray*}
where the phase of the tachyon vertex operator is 
$$
-i(\Theta,\psi_{\pm}^{\zeta^{\psi}})^{\dagger}_{u}=\int_{0}^{2\pi}\frac{d\sigma}{2\pi}\int_{0}^{2\pi}\frac{d\sigma^{\prime}}{2\pi}\hspace{1mm}(\psi_{\pm}^{\zeta^{\psi}})^{\dagger}(\sigma^{\prime})\hspace{0.5mm}(\Delta_{u}^{NSNS})^{*}(\sigma,\sigma^{\prime})\hspace{0.5mm}\Theta^{\dagger}(\sigma).
$$
The Majorana condition of $\psi$ is satisfied by
$$ \psi_{r}^{\dagger}=\psi_{-r}, \quad \widetilde{\psi}_{r}^{\dagger}=-\widetilde{\psi}_{-r}  \hspace{5mm}(r\in\mathbf{Z}\!+\!1/2).$$
We further require that the eigenvalue of $\psi^{\zeta^{\psi}_{o}}$ on $|T_{u,\eta}^{\zeta^{\psi}}(\Theta)\rangle$ and that of $(\psi^{\zeta^{\psi}_{o}})^{\dagger}$ on $\langle T_{u,\eta}^{\zeta^{\psi}}(\Theta)|$ coincide. From this we have
$$ \theta_{r}^\dagger=\bar{\theta}_{r}, \qquad \bar{\theta}_{r}^\dagger=\theta_{r} \hspace{8mm}(r\in\mathbf{Z}_{\geq 0}\!+\!1/2).$$ 
They are sufficient to guarantee that the eigenvalue function $\xi^\psi_c(\sigma)$ itself is a Majorana field. \\[1mm]

We are now ready to calculate (\ref{<psi>def}) on the well-defined tachyon states. Using the {\tt [Formula\hspace{0.7mm}2]} in Appendix A, 
$$ f_{r}:=-i\eta^{\prime}\!\rho, \qquad  g_{r}:=i\eta\rho e^{-2rs}, $$
$$ a_{r}:=-c_{r}^{v}\theta_{r}, \qquad \bar{a}_{r}:=-i\zeta^{\psi} c_{r}^{v}\bar{\theta}_{r}, \qquad b_{r}:=-c_{r}^{u}\bar{\theta}_{r}e^{-rs}, \qquad \bar{b}_{r}:=-i\zeta^{\psi}c_{r}^{u}\theta_{r}e^{-rs}, $$
we find
\begin{eqnarray}
\langle \psi^{\zeta_{o}^{\psi}}(\sigma)\rangle^{\zeta^{\psi}}
&=&\lim_{s \to 0}\pr2 \frac{{}_{f_{r}} \langle a_{r},\bar{a}_{r}|\psi^{\zeta_{o}^{\psi}}(\sigma)|b_{r},\bar{b}_{r} \rangle_{g_{r}}^{\zeta^{\psi}}}{{}_{f_{r}} \langle a_{r},\bar{a}_{r}|b_{r},\bar{b}_{r}\rangle_{g_{r}}^{\zeta^{\psi}}} \nonumber \\[2mm]
&=&\frac{1-\zeta^{\psi}\zeta_{o}^{\psi}}{2}\xi_{c}^{\psi}(\sigma).  
\label{<psi>}
\end{eqnarray}
From this formula we see that $\xi_{c}^{\psi}(\sigma)$  is the fermionic D-brane coordinate in the T-dual space.
Note that the result (\ref{<psi>}) is independent of the sign of $\eta\eta^{\prime}\!\rho\rho^{\prime}$. 

Moreover, if $\zeta^{\psi}\zeta^{\psi}_{o}=\zeta\zeta_{o}$, combining (\ref{<X>}) and (\ref{<psi>}) together, we obtain an expectation value of supersymmetric D-brane coordinate $\mathbf{X}=X+i\epsilon\psi$ as
\begin{equation}
\langle \mathbf{X}^{\zeta_{o}}(\sigma)\rangle^{\zeta}
=\frac{1-\zeta\zeta_{o}}{2}(\xi^{X}_{c}(\sigma)+i\epsilon\xi^{\psi}_{c}(\sigma))
=\frac{1-\zeta\zeta_{o}}{2}\xi^{\mathbf{X}}_{c}(\sigma),
\end{equation}
where we introduced a constant Grassmann-odd number $\epsilon$.
\section{Tachyon correlation functions in one dimension}

The tachyon states $|T\rangle$ were defined in \S II to provide local information of D-branes. In fact it was shown in \S III they are eigenstates of the string coordinate on the D-brane whose eigenvalues can be identified with soliton coordinates of the KP systems. In the rest of this paper we are going to study properties of correlation functions of these states. In this section we investigate local properties of tachyon correlators of bosonic part in \S IV-A and the fermionic one in \S IV-B in one dimension as the preparation for the next section. In \S IV-C we integrate these correlators over the tachyon momenta to see their global features how these amplitudes are affected through the tachyon condensation depending on the on-shell values. Correspondingly we classify the decay of unstable D-branes, which is known as the \textit{descent relations} \cite{Sen}.
\subsection{Bosonic Part}
Consider the following bosonic correlators
$$ \langle \rho^{\prime}|:e^{-i(K',X^{\zeta^{\prime}})^{\dagger}_{v}}:e^{-s(L_{0}+\widetilde{L}_{0})}:e^{i(K,X^{\zeta})_{u}}:|\rho\rangle\hspace{0.5mm}.$$
Recall that, as is seen in (\ref{b-state}), the tachyon vertex operator $:e^{i(K,X^\zeta)_{u}}:$ affects the boundary state $|\rho\rangle$ only when $\zeta\rho=-1$ 
$$
 :e^{i(K,X^{\zeta})_{u}}:|\rho\rangle \\[2mm]=\left\{
 \begin{array}{ll}|T_{u}^{\zeta}(K)\rangle &\hspace{6.4mm}(\zeta\rho=-1) \cr |\rho\rangle &\hspace{6.4mm} (\zeta\rho=+1)\cr
\end{array}.\right.
$$
We will call the correlators calculated by the tachyon ($\zeta\rho=-1$) and the non-tachyon ($\zeta\rho=1$) states \textit{tachyon correlation functions}. Then there are four types of correlators because they are classified by the two independent cases of $\zeta\rho=\pm 1$, $\zeta^{\prime}\rho^{\prime}=\pm 1$. For later convenience we label them as $B_{1}\sim B_{4}$ ($B$ denotes Bosonic part). 
From {\tt [Formula\hspace{0.7mm}1]} in Appendix A, these correlators are calculated as follows:

\begin{eqnarray}
B_{1}
&:=&\langle T_{v}^{\zeta}(K')|e^{-s(L_{0}+\widetilde{L}_{0})}|T_{u}^{\zeta}(K)\rangle \nonumber \\[2mm]
&=&
\prod_{n=1}^{\infty}\frac{1}{1-\rho\rho' e^{-2ns}}\exp\Bigg[{(c_{n}^{u})^{2}p_n\bar p_n+(c_{n}^{v})^{2}p'_n\bar p'_n-c_{n}^{u}c_{n}^{v}e^{ns}(p_n\bar p'_n+\rho\rho'p'_n\bar p_n)
\over n(1-\rho\rho' e^{2ns})}
\Bigg] \label{B_{1}} \hspace{0.5mm}, \\[2mm]
B_{2}&:=&\langle \rho^{\prime}|e^{-s(L_{0}+\widetilde{L}_{0})}|\rho\rangle=\prod_{n=1}^{\infty}\frac{1}{1-\rho\rho' e^{-2ns}} \hspace{0.5mm}, \nonumber \\[2mm]
B_{3}
&:=&\langle \rho^{\prime}|e^{-s(L_{0}+\widetilde{L}_{0})}|T_{u}^{\zeta}(K) \rangle 
=
\prod_{n=1}^{\infty}\frac{1}{1-\rho\rho' e^{-2ns}}\exp\Bigg[{(c_{n}^{u})^{2}p_n\bar p_n
\over n(1-\rho\rho' e^{2ns})}
\Bigg] \nonumber\hspace{0.5mm}, \\[2mm]
B_{4}
&:=&\langle T_{v}^{\zeta'}(K')|e^{-s(L_{0}+\widetilde{L}_{0})}|\rho\rangle 
=
\prod_{n=1}^{\infty}\frac{1}{1-\rho\rho' e^{-2ns}}\exp\Bigg[{(c_{n}^{v})^{2}p'_n\bar p'_n
\over n(1-\rho\rho' e^{2ns})}
\Bigg] . \nonumber 
\end{eqnarray}

We recognize easily the following facts.
\begin{itemize}
\item
Starting from $B_1$ the other correlators $B_3,\ B_4$ and $B_2$ can be obtained by taking the limits of either $c^v_n\rightarrow 0$ or $c^u_n\rightarrow 0$ or both.
\item
The momentum distribution is a Gaussian, as we expect from our construction. The full condensation of tachyons takes place when $c_n^u=1$ for all $n$.
\item
In the case of $v=u$ and $\rho=\rho'$, $B_1$ has the following expression
\begin{eqnarray*}
B_1
=
\prod_{n=1}^\infty{1\over 1-e^{-2ns}}\exp\left[{(c_n^u)^2\over n}\left(
{p_n\bar p_n+p'_n\bar p'_n\over 1+e^{ns}}
-{(p_n-p'_n)(\bar p_n-\bar p'_n)\over e^{ns}-e^{-ns}}\right)\right]
\end{eqnarray*}
When the distance $s$ between the D-branes is large the first term in the exponent dominates the behavior of the correlator which represents two independent D-branes. As the distance becomes small the second term, which represents the interference between the D-branes, becomes large and turns to 
$$
B_1\ \mathop{\longrightarrow}_{s\rightarrow 0}\ \prod_{n=1}^\infty{2n\pi\over i(c_n^u)^2}e^{(c_n^u)^2p_n\bar p_n/n}\delta(p_n-p'_n)\delta(\bar p_n-\bar p'_n)
$$
when the D-branes clash.
\end{itemize}

\subsection{Fermionic Part}
Consider the following fermionic correlators
$$
\langle \eta^{\prime},\rho^{\prime}|:e^{-i(\Theta',\psi^{\zeta^{\prime \psi}})_{v}^{\dagger}}:e^{-s(L_{0}+\widetilde{L}_{0})}:e^{i(\Theta,\psi^{\zeta^{\psi}})_{u}}:|\eta,\rho\rangle\hspace{0.5mm}.
$$
The fermionic tachyon state $|T_{u}^{\zeta^\psi}(\Theta)\rangle$ and non tachyon state $|\eta,\rho\rangle$ correspond to the following combinations of indices
$$
 :e^{i(\Theta,\psi^{\zeta^\psi})_{u}}:|\eta,\rho\rangle \\[2mm]=\left\{
 \begin{array}{ll}|T_{u,\eta}^{\zeta^\psi}(\Theta)\rangle &\hspace{6.4mm}(\zeta^{\psi}\eta\rho=1) \cr |\eta,\rho\rangle &\hspace{6.4mm} (\zeta^{\psi}\eta\rho=-1)\cr
\end{array}.\right.
$$
There are eight types of correlators because they are classified by the two independent cases of $\zeta^{\psi}\eta\rho=\pm 1$, $\zeta^{\prime \psi}\eta^{\prime}\rho^{\prime}=\pm 1$. 
They have the independent spin structures on both sides $(\eta^{\prime},\eta)$, and depend only on their products $\eta\!\cdot\!\eta^{\prime}=\pm 1.$
So we label these correlators as $F_{1}^{\hspace{0.7mm}\eta\cdot\eta^{\prime}} \!\sim F_{4}^{\hspace{0.7mm}\eta\cdot\eta^{\prime}}$ ($F$ denotes Fermionic part). 
From {\tt [Formula\hspace{0.7mm}1]} in Appendix A, these correlators are calculated as follows:
\begin{eqnarray}
F_{1}^{\eta\cdot\eta^{\prime}}
&:=&
\langle T_{v,\eta^{\prime}}^{\zeta^{\prime \psi}}(\Theta')|e^{-s(L_{0}+\widetilde{L}_{0})}|T_{u,\eta}^{\zeta^{\psi}}(\Theta)\rangle  \nonumber \\[2mm]
&=&
\prod_{r=\frac{1}{2}}^{\infty}(1-\eta\eta^{\prime}\rho\rho'e^{-2rs})\exp\!\bigg[{(c^u_r)^2\theta_r\bar\theta_r+(c^v_r)^2\theta'_r\bar\theta'_r-c^u_rc^v_re^{rs}(\theta_r\bar\theta'_r+\eta\eta'\rho\rho'\theta'_r\bar\theta_r)
\over 1-\eta\eta'\rho\rho'e^{2rs}}\bigg] \hspace{0.5mm}, \\[3mm]
F_{2}^{\eta\cdot\eta^{\prime}}
&:=&\langle \eta^{\prime},\rho^{\prime}|e^{-s(L_{0}+\widetilde{L}_{0})}|\eta,\rho\rangle
=\pr2(1-\eta\eta^{\prime}\rho\rho'e^{-2rs})\hspace{0.5mm}, \nonumber \\[2mm]
F_{3}^{\eta\cdot\eta^{\prime}}
&:=&\langle \eta^{\prime},\rho^{\prime}|e^{-s(L_{0}+\widetilde{L}_{0})}|T_{u,\eta}^{\zeta^{\psi}}(\Theta)\rangle
=\prod_{r=\frac{1}{2}}^{\infty}(1-\eta\eta^{\prime}\rho\rho'e^{-2rs})\exp\!\bigg[\frac{(c_{r}^{u})^{2}\theta_{r}\bar{\theta}_{r}}{1-\eta\eta^{\prime}\rho\rho'e^{2rs}}\bigg] \hspace{0.5mm}, \\[3mm]
F_{4}^{\eta\cdot\eta^{\prime}}
&:=&\langle T_{v,\eta^{\prime}}^{\zeta^{\prime \psi}}(\Theta')|e^{-s(L_{0}+\widetilde{L}_{0})}|\eta,\rho\rangle
=\prod_{r=\frac{1}{2}}^{\infty}(1-\eta\eta^{\prime}\rho\rho'e^{-2rs})\exp\!\bigg[\frac{(c_{r}^{v})^{2}\theta'_{r}\bar{\theta'}_{r}}{1-\eta\eta^{\prime}\rho\rho'e^{2rs}}\bigg] \hspace{0.5mm}. \\[3mm]
\nonumber
\end{eqnarray}

We observe immediately that
\begin{itemize}
\item
The supersymmetry is manifest between $B_j$ and $F_j$. \item
$F_3,\ F_4,\ F_2$ are obtained starting from $F_1$ by reductions of $c_r^v\rightarrow 0$, $c_r^u\rightarrow 0$ and both, respectively.
\item
If we specify $F_1$ by $u=v$ and $\eta\rho=\eta'\rho'$\,\,$(\textit{i.e.},\,\,\zeta^{\psi}=\zeta^{\prime \psi})$, we obtain
$$
F_1=\prod_{r=\frac{1}{2}}^{\infty}(1-e^{-2rs})\exp\left[(c^u_r)^2\left({\theta_r\bar\theta_r+\theta'_r\bar\theta'_r\over 1+e^{rs}}-{(\theta_r-\theta'_r)(\bar\theta_r-\bar\theta'_r)\over e^{rs}-e^{-rs}}\right)\right].
$$
As it was the case of $B_1$ the first term in the exponent represents the contribution from the two independent D-branes. The second term shows their interference, which turns to
$$
F_1\ \mathop{\longrightarrow}_{s\rightarrow 0}\ \prod_{r={1\over 2}}^\infty{(c_r^u)^2}e^{(c_r^u)^2\theta_r\bar\theta_r}\delta(\theta_r-\theta'_r)\delta(\bar \theta_r-\bar\theta'_r)
$$
when they clash.
\end{itemize}
\subsection{On-Shell Correlators}
Having calculated the tachyon correlators in each sector (\S IV-A, B), we study in this subsection their global properties by integrating them over the tachyon momenta.
We here demonstrate only the $B_{1}\!\cdot\!F_{1}^{\eta\cdot\eta^{\prime}}$ case, which reduces to the other cases.

Integrating the correlator $B_{1}\!\cdot\!F_{1}^{\eta\cdot\eta^{\prime}}$ over the tachyon momenta $K$, $K'$, $\Theta$, $\Theta'$,\hspace{1mm} we obtain 
\begin{eqnarray}
& &{}_{NSNS}\langle\eta',\rho'|\otimes\langle \rho'|\Phi^\dagger_v(X,\psi)e^{-s(L_0+\tilde L_0)}\Phi_u(X,\psi)|\rho\rangle\otimes|\eta,\rho\rangle_{NSNS} \nonumber \\[2mm]
& &\quad = \prod_{n=1}^\infty\prod_{r=\frac{1}{2}}^{\infty}\left({c_{n}^{u}c_{n}^{v}\over c_{r}^{u}c_{r}^{v}}\right)^{2}{1-\eta\eta'\rho\rho' (1-2(c_r^u)^2)(1-2(c_r^v)^2)e^{-2rs}\over 1-\rho\rho'(1-2(c_n^u)^2)(1-2(c_n^v)^2)e^{-2ns}}
\label{correlators after path integral}
\end{eqnarray}
where $\Phi_u(X,\psi)$ is the full tachyon field defined by (\ref{full tachyon field}).

 We then examine how they behave according to the on-shell values: $(u,v)=(0,0)$, $(0, \infty)$, $(\infty, 0)$, $(\infty, \infty)$. So far we have not specified the values of $c_{n}$'s, $c_{r}$'s within this section and note that (\ref{correlators after path integral}) is the general amplitude as the function of $c_{n,r}$'s. Now we specify them as de Alwis type's (\ref{c_n^u}) (\ref{c_r^u}) and obtain the on-shell correlators as
\begin{eqnarray}
& &(u,v)=(0,0)\hspace{2mm};\hspace{3mm}\langle \Phi\rangle^{-1+\frac{\rho+\rho^{\prime}}{2}}\prod_{n=1}^\infty\prod_{r=\frac{1}{2}}^{\infty}\frac{1-\eta\eta^{\prime}e^{-2rs}}{1-e^{-2ns}}, \nonumber \\[2mm]
& &(u,v)=\cases{(\infty,0)\cr (0,\infty)\cr}\hspace{2mm};\hspace{3mm}\langle \Phi\rangle^{\frac{\rho+\rho^{\prime}}{2}}\prod_{n=1}^\infty\prod_{r=\frac{1}{2}}^{\infty}\frac{1+\eta\eta^{\prime}e^{-2rs}}{1+e^{-2ns}}, 
\label{on-shell correrators of de Alwis type} \\[2mm]
& &(u,v)=(\infty,\infty)\hspace{2mm};\hspace{3mm}\langle \Phi\rangle^{1+\frac{\rho+\rho^{\prime}}{2}}\prod_{n=1}^\infty\prod_{r=\frac{1}{2}}^{\infty}\frac{1-\eta\eta^{\prime}e^{-2rs}}{1-e^{-2ns}}, \nonumber
\end{eqnarray}
where $\langle\Phi\rangle$ is the vev of the tachyon field $\Phi_{u,\rho}$ in the case of $(u,\rho)=(\infty,+1)$, and given by
\begin{equation}
\langle\Phi\rangle := \prod_{n=1}^{\infty}\prod_{r=\frac{1}{2}}^{\infty}\left(\frac{n}{r}\right).
\label{vev of tachyon field}
\end{equation} 
The infinite product is properly regularized as $\sqrt{\pi}$, which we will derive in Appendix B. It is found from (\ref{on-shell correrators of de Alwis type}) that the factor $\langle\Phi\rangle^{m}$\,\,$(-2\!\leq\!m\!\leq\!2,\,\,\,m\!\in\!\mathbf{Z})$ appears in front of the usual amplitudes. This indicates that, if $m \neq 0$, the tachyons infinitely condense and the conformal invariance is restored.
At the same time the decay of unstable D-branes should occur. The decay types are classified by the absolute value of $m$ and there are two types (\,$|m|=1,\,2$\,). The classification of the decay of unstable D-branes is known as the \textit{descent relations} \cite{Sen}. 

Let us see, for example, the decay types in the case of $(\rho,\rho^{\prime})=(+1,+1)$ \cite{TST1}. There are the three cases where the tachyon condensation indeed occurs: $(u,v)=(0, \infty)$, $(\infty, 0)$, $(\infty, \infty)$.
One corresponds to the \textit{kink}-type configuration ($m=1$), whose codimension is one, and the descent relation from a non-BPS $p$-brane to a BPS $(p-1)$-brane:
$$ (u,v)=\cases{(\infty,0)\cr (0,\infty)\cr}\hspace{1mm};\hspace{3mm}\langle\Phi\rangle\prod_{n=1}^{\infty}\prod_{r=\frac{1}{2}}^{\infty}\frac{1+\eta\eta^{\prime}e^{-2rs}}{1+e^{-2ns}}. $$
The other corresponds to the \textit{vortex}-type configuration ($m=2$), whose codimension is two, and the descent relation from a non-BPS $p$-brane to a BPS $(p-2)$-brane:
$$ (u,v)=(\infty,\infty)\hspace{1mm};\hspace{3mm}\langle\Phi\rangle^{2}\prod_{n=1}^{\infty}\prod_{r=\frac{1}{2}}^{\infty}\frac{1-\eta\eta^{\prime}e^{-2rs}}{1-e^{-2ns}}. $$

The other $(\rho,\rho^{\prime})$ cases are similar but different only from the appearace of the inverse of $\langle\Phi\rangle$. We call these inverse types $\langle\Phi\rangle^{-1}$, $\langle\Phi\rangle^{-2}$ \textit{anti-kink}, \textit{anti-vortex} respectively. Thus the descent relations are summarized as follows.
\begin{center}
 \begin{tabular}{|c||c|c|c|}\hline
   $(\rho,\rho^{\prime})$    & $(u,v)=(\infty,\infty)$ & $(u,v)=(\infty,0)$, $(0,\infty)$ & $(u,v)=(0,0)$ \\ \hline
   $(+1,+1)$  &  vortex $\langle\Phi\rangle^{2}$ & kink $\langle\Phi\rangle^{1}$ & \,\,no condensation $\langle\Phi\rangle^{0}$\,\, \\ \hline
   $(\pm 1,\mp 1)$ & kink $\langle\Phi\rangle^{1}$ & \,\,no condensation $\langle\Phi\rangle^{0}$\,\,& anti-kink $\langle\Phi\rangle^{-1}$ \\ \hline
   $(-1,-1)$  & \,\,no condensation $\langle\Phi\rangle^{0}$\,\,& anti-kink $\langle\Phi\rangle^{-1}$ & anti-vortex $\langle\Phi\rangle^{-2}$ \\ \hline
 \end{tabular}
\end{center}

We have seen in this subsection the followings. 
\begin{itemize}
\item As a token of the tachyon condensation, $\langle\Phi\rangle$ (\,eq.(\ref{vev of tachyon field})\,) appears in front of the usual amplitudes, 
\item The decay types of unstable D-branes are classified by the power $m$\,(\,$|m|=1,2$\,) of $\langle\Phi\rangle$. 
\end{itemize}  
\section{Tachyon correlation functions in ten dimensions}
In this section we consider the physics in the ten-dimensional space-time by using the results in the previous section. Our plan in this section is as follows.
First we evaluate the tachyon correlation functions in ten dimensions, which always have both tachyon and non-tachyon states for either the Neumann or the Dirichlet directions.
Second we integrate these correlators over the tachyon momenta and take the parameters $(u,v)$ as the on-shell values to see how the amplitudes are affected through the insertion of the tachyon fields into the original amplitude, {\it i.e.}, the tachyon condensation.
Throughtout this section we take the light cone gauge and $\mu=8,9$ as its directions so that we ignore the ghost sector (the double Wick rotated formalism as in \cite{BG}),
and restrict the parameters between the bosonic and the fermionic field to $\zeta=-\zeta^{\psi}\!\cdot\eta$ since $\rho$'s have in common. \\[1mm] 

We first define the NSNS sector of $D_{p}$-brane boundary state, which is appropriately GSO-projected out
\begin{equation}
|D_{p}\rangle_{NSNS}=\frac{1}{\sqrt{2}}(\hspace{1.3mm}|D_{p},\eta=+1\rangle_{NSNS}-|D_{p},\eta=-1\rangle_{NSNS}\hspace{1mm})\hspace{0.6mm},
\label{|D_p>}
\end{equation}
where
\begin{equation}
|D_{p},\eta\rangle_{NSNS}={\mathcal N}_{p}\hspace{0.7mm}\exp\!\bigg[\sum_{n=1}^{\infty}\frac{1}{n}\alpha_{-n}^{\mu}S_{\mu\nu}\widetilde{\alpha}_{-n}^{\nu}+i\eta\sum_{r=\frac{1}{2}}^{\infty}\psi_{-r}^{\mu}S_{\mu\nu}\widetilde{\psi}_{-r}^{\nu}\bigg]|0\rangle\otimes |0\rangle_{NSNS}\hspace{0.5mm}.
\label{|D_p eta>}
\end{equation}
$S_{\mu\nu}$ is a diagonal $8\times 8$ matrix with $\rho=-1$ for the Neumann ($\mu=0,\cdots,p$) and $\rho=+1$ for the Dirichlet ($\mu=p+1,\cdots,7$) directions. 
The normalization constant ${\mathcal N}_{p}$ is determined by Cardy's condition (open/closed duality) \cite{C}:
$$
{\mathcal N}_{p}=\frac{\sqrt{\pi}(2\pi\sqrt{\alpha^{\prime}})^{3-p}}{2}\left(=\frac{\kappa T_{p}}{2}\right),
$$
where $\kappa\hspace{0.5mm}(\hspace{0.5mm}=\hspace{0.5mm}8\pi^{7/2}\alpha^{\prime \hspace{0.4mm} 2}g_{st})$ is the ten-dimensional gravitational constant and $T_{p}$ is the $D_{p}$-brane tension given by
\begin{equation}
T_{p}=\frac{1}{(2\pi)^{p}\hspace{1mm}(\alpha^{\prime})^{\frac{p+1}{2}}\hspace{0.7mm}g_{st}},
\label{T_p}
\end{equation}
and $g_{st}$ is the coupling constant in string theory. This is obtained as follows. $g_{st}$ is defined as the ratio of tensions between F-string and D-string ($=$ D1-brane):
$$ g_{st} \equiv \frac{T_{F1}}{T_{1}}=\frac{1}{2\pi\alpha^{\prime}T_{1}}, $$
and we have $T_{1}=g_{st}^{-1}\frac{1}{2\pi\alpha^{\prime}}$. This and the ratio between D-brane tensions (\ref{D-brane tensions ratio}) give us (\ref{T_p}). \\[1mm]

From (\ref{|D_p>}) and (\ref{|D_p eta>}) we obtain the original amplitude for the NSNS sector as 
\begin{equation}
{}_{NSNS} \langle D_{p}|e^{-s(L_{0}+\widetilde{L}_{0}-1)}|D_{p}\rangle_{NSNS}=|{\mathcal N}_{p}|^{2}\hspace{0.5mm}e^{s}\prod_{n=1}^{\infty}\prod_{r=\frac{1}{2}}^{\infty}\bigg[\bigg(\frac{1+e^{-2rs}}{1-e^{-2ns}}\bigg)^{\! 8}-\bigg(\frac{1-e^{-2rs}}{1-e^{-2ns}}\bigg)^{\! 8}\hspace{1mm}\bigg].
\label{original amplitude}
\end{equation}
We then insert the tachyon fields into this amplitude in order to see the effect of the tachyon condensation: 
\begin{eqnarray}
& &{}_{NSNS} \langle D_{p}|\Phi^{\dagger}_{v}(X,\psi)e^{-s(L_{0}+\widetilde{L}_{0}-1)}\Phi_{u}(X,\psi)|D_{p}\rangle_{NSNS} 
\label{insertion} \\[2mm]
&\propto&\int{\mathcal D}K\tilde{\Phi}(K)\int{\mathcal D}K^{\dagger}\tilde{\Phi}^{\dagger}(K)\int{\mathcal D}\Theta\tilde{\Phi}(\Theta)\int{\mathcal D}\Theta^{\dagger}\tilde{\Phi}^{\dagger}(\Theta) \nonumber \\[2mm]
& &\times {}_{NSNS} \langle D_{p}|:e^{-i(K,X^{\zeta^{\prime}})^{\dagger}_{v}-i(\Theta,\psi^{-\zeta^{\prime}\cdot\eta^{\prime}})^{\dagger}_{v}}:e^{s(L_{0}+\widetilde{L}_{0})}:e^{i(K,X^{\zeta})_{u}+i(\Theta,\psi^{-\zeta\cdot\eta})_{u}}:|D_{p}\rangle_{NSNS}.
\nonumber 
\end{eqnarray}
Recall that the tachyon vertex serves as a projection operator, {\it i.e.}, alternatively chooses tachyon or non-tachyon states according to $\zeta$:
$$ :e^{i(K,X^{\zeta})_{u}+i(\Theta,\psi^{-\zeta\cdot\eta})_{u}}:|D_{p}\rangle_{NSNS}=\cases{|T_{u,\eta}(K,\Theta)\rangle^{\parallel} \otimes |\eta,\rho=+1\rangle^{\bot} \hspace{5mm}(\zeta=+1) \cr|\eta,\rho=-1 \rangle^{\parallel} \otimes |T_{u,\eta}(K,\Theta)\rangle^{\bot} \hspace{5mm}(\zeta=-1),} $$
where $\parallel$ and $\bot$ denote $\mu=0,\cdots, p$ and $\mu=p+1,\cdots,7$. Let us now evaluate the integrands of (\ref{insertion}). From the results in the previous section they are obtained as follows: 
\begin{eqnarray}
& &\bullet\hspace{2mm}\zeta=+1\hspace{2mm}(ordinary)\hspace{1mm};\hspace{5mm}(B_{2})^{p+1}(B_{1})^{7-p}\bigg[(F_{2}^{\hspace{0.7mm}+})^{p+1}(F_{1}^{\hspace{0.7mm}+})^{7-p}-(F_{2}^{\hspace{0.7mm}-})^{p+1}(F_{1}^{\hspace{0.7mm}-})^{7-p}\bigg],  \\[2mm]
& &\bullet\hspace{2mm}\zeta=-1\hspace{2mm}(\hspace{0.3mm}T\!-\!dual\hspace{0.3mm})\hspace{1.5mm};\hspace{5.5mm}(B_{1})^{p+1}(B_{2})^{7-p}\bigg[(F_{1}^{\hspace{0.7mm}+})^{p+1}(F_{2}^{\hspace{0.7mm}+})^{7-p}-(F_{1}^{\hspace{0.7mm}-})^{p+1}(F_{2}^{\hspace{0.7mm}-})^{7-p}\bigg].
\end{eqnarray}
It is found that taking the T-duality is equivalent to exchanging each other for the three pairs $B_{1}\leftrightarrow B_{2}$, $F_{1}^{\hspace{0.7mm}\pm}\leftrightarrow F_{2}^{\hspace{0.7mm}\pm}$ and in particular both are the same if $p=3$: 
$$
\hspace{0.5mm}(B_{1}B_{2})^{4}\bigg[(F_{1}^{\hspace{0.7mm}+}F_{2}^{\hspace{0.7mm}+})^{4}-(F_{1}^{\hspace{0.7mm}-}F_{2}^{\hspace{0.7mm}-})^{4}\bigg]. \\
$$

Here we comment on the T-duality as mentioned in \S II.\hspace*{1mm} In the $\zeta=+1$ case, the tachyon state can be generated from only the Neumann directions, which means that the effect of the tachyons can be seen from only the directions. On the other hand, the Dirichlet directions are the on-shell ones because no tachyons exist in the directions. 
\begin{eqnarray*}
0 \leftarrow &u& \rightarrow \infty \nonumber \\ \zeta=+1\hspace{1mm};\hspace{7mm} 
0 \leftarrow &c_{n,r}^{u}& \rightarrow 1 \\ |N\rangle \leftarrow &|B\rangle& \rightarrow |D\rangle \nonumber
\end{eqnarray*}
In the $\zeta=-1$ case, the above discussion is reversed about the Neumann and the Dirichlet directions.
\begin{eqnarray*}
0 \leftarrow &u& \rightarrow \infty \nonumber \\ \zeta=-1\hspace{1mm};\hspace{7mm}
1 \leftarrow &c_{n,r}^{u}& \rightarrow 0 \\ |N\rangle \leftarrow &|B\rangle& \rightarrow |D\rangle \nonumber 
\end{eqnarray*}

Next we integrate these correlators over the tachyon momenta and take the on-shell limit within $\zeta=-1$. 
We take the parameters $(u,v)$ as the two types of the on-shell values as well as the previous section. One amplitude corresponds to the kink-type configuration:
\begin{equation}
(u,v)=\cases{(\infty,0)\cr (0,\infty)\cr}\hspace{0.5mm};\hspace{1mm}|{\mathcal N}_{p}|^{2}\hspace{0.5mm}e^{s}\hspace{0.5mm}\langle\Phi\rangle^{7-p}\prod_{n=1}^{\infty}\prod_{r=\frac{1}{2}}^{\infty}\bigg[\bigg(\frac{1+e^{-2rs}}{1+e^{-2ns}}\bigg)^{\! p+1}\!\bigg(\frac{1-e^{-2rs}}{1-e^{-2ns}}\bigg)^{\! 7-p}\!\!\!-\hspace{0.5mm}\bigg(\frac{1-e^{-2rs}}{1+e^{-2ns}}\bigg)^{\! p+1}\!\bigg(\frac{1+e^{-2rs}}{1-e^{-2ns}}\bigg)^{\! 7-p}\hspace{1mm}\bigg].
\end{equation}
This amplitude vanishes only when $p=3$ and this indicates that, according to the descent relation, the non-BPS $D3$-brane decays to the BPS $D2$-brane through the tachyon condensation. The other corresponds to the vortex-type configuration: 
\begin{equation}
(u,v)=(\infty,\infty)\hspace{2mm};\hspace{3mm}|{\mathcal N}_{p}|^{2}\hspace{0.5mm}e^{s}\hspace{0.5mm}\langle\Phi\rangle^{2(7-p)}\prod_{n=1}^{\infty}\prod_{r=\frac{1}{2}}^{\infty}\bigg[\bigg(\frac{1-e^{-2rs}}{1-e^{-2ns}}\bigg)^{\! 8}-\bigg(\frac{1+e^{-2rs}}{1-e^{-2ns}}\bigg)^{\! 8}\hspace{1mm}\bigg]\hspace{1mm}.
\end{equation}
\section{Summary and remarks}

We introduced a tachyon field (\ref{tachyon state as our starting point}), (\ref{tachyon field for fermionic part}), which simply connected to the calculus of the tachyon condensation. In \S II we constructed it by integrating the generalized vertex operator over the tachyon momenta. This field acts on the bare boundary state (\ref{bare boundary state}), (\ref{bare NSNS boundary state}): either the Neumann or the Dirichlet state, and generates the boundary state suggested by S.P. de Alwis under the conditions of $\zeta\rho=-1$ for the bosonic, and $\zeta^{\psi}\eta\rho=+1$ for the fermionic fields. Note that our formulation essentially owes to the nature of the bare boundary state.

On the other hand we have mainly studied string theory within the framework of integrable systems. In particular we have noted the Hirota-Miwa equation (\ref{original HBDE}), which characterizes the KP hierarchy in integrable systems. In \S III-A we generalized it to the case where there are two boundaries and showed that correlation functions of local tachyon fields satisfied the generalized Hirota-Miwa equation (\ref{bilinear eq with boundary}). Using the formalism based on this evidence, we demonstrated in \S III-B,C that the soliton coordinates $\xi$ in integrable systems could be identified with the expectation values of the space-time coordinates $\xi_c$ on an unstable D-brane, where tachyon fields were attached. They appear as the space-time coordinates in the T-dual theory. The observation in \S III is the most important in this paper and provides an explicit correspondence between string theory and integrable systems.

We also evaluated possible tachyon correlation functions to obtain the local information about tachyons on an unstable D-brane. In \S IV-A,(B) we presented some types of correlators corresponding to various combinations of $\zeta\rho=\pm 1$ and $\zeta^{\psi}\eta\rho=\pm 1$, and gave the local picture of the full condensation of tachyons under $c_{n,r}=1$. In \S IV-C, by integrating the tachyon correlation functions over the tachyon momenta and taking the on-shell limit, we saw how these amplitudes were affected through the insertion of the tachyon field into the original amplitudes, {\it i.e.}, the tachyon condensation. It is found that the vev of the tachyon field $\langle \Phi\rangle$ in the middle of the condensation appears and this explains the descent relation, which is the classification of the decay of unstable D-branes.

Finally in \S V we applied the one-dimensional results in the previous section to the ten-dimensional case. The tachyon field inserted into the original amplitude (\ref{original amplitude}) serves as a projection operator for the boundary state (\ref{|D_p>}) and generates various amplitudes. To get more results we need further calculations in detail, which we will report elsewhere.

\vglue 0.5cm
We would like to close the paper with some remarks.

The tachyon states we defined do depend on the parameter $\zeta$, which can take values either $+1$ or $-1$ corresponding to the ordinary theory and the T-dual theory. This enables us to study these two theories in an equal footing. The key observation based on this formalism is that the tachyon state thus defined plays the role of a projection operator as presented in (\ref{b-state}), (\ref{fermionic relation}). We see clearly that such property depnends not on the signs of $\zeta$ and $\rho$ separately, but on their product $\zeta\rho$. This is also true in the fermionic sector, which depends on the particular combination $\zeta^\psi\eta\rho$.

This property of the tachyon states determines the nature of the fields which are diagonal in these states. As we have found in \S III-B,C the bosonic coordinate $X^{\zeta_o}$ is diagonal when $\zeta_o\rho=+1$ while the fermionic coordinate $\psi^{\zeta_o^{\psi}}$ is diagonal when $\zeta_o^{\psi}\eta\rho=-1$. When $\rho$ and $\eta$ are fixed, the space-time coordinates which are diagonal are those of T-dual theory. This is what we could expect from the nature of duality. What is highly nontrivial is that the `dual coordinate' is associated with a soliton coordinate. If the free energy of the system is localized in the soliton coordinate, for example, it means the localization of the D-brane in the `dual coordinate'. Moreover this fact provides an interpretation of the Miwa transformation. Namely the Miwa transformation connects the momenta $k_j$'s in the ordinary space-time with the `dual coordinate' $\xi_c$.

Final remark owes to the fact that the eigenvalues $\xi_c$ do not distinguish the values of $\zeta_o$. The theories of the same value of $\zeta_o\rho$ or $\zeta_o^{\psi}\eta\rho$ are equivalent. In other words the soliton coordinates $\xi$, hence the Hirota-Miwa equation itself, describe the dynamics of D-branes of both ordinary and dual theories simultaneously. From this point of view the Hirota-Miwa equation manifests the T-duality, and the distinction of the theories arises only in its solutions.
\\[5mm]
{\bf Acknowledgement}

The authors would like to thank Mr. Hironori Yamaguchi for discussions. 

{\bf Appendix A}

We present the following two formulae. Each formula has the parallel form between bosonic and fermionic part.
Using these formulae, we can easily calculate the tachyon correlation functions and the expectation values on tachyon states.
{\tt [Formula\hspace{0.7mm}1]} is used for the tachyon correlation functions in \S IV and {\tt [Formula\hspace{0.7mm}2]}, 
which is a corollary of {\tt [Formula\hspace{0.7mm}1]}, is used for the expectation values on tachyon states in \S III-B, C. \\[6mm]
{\bf [Bosonic part]} \\[2mm]
$a_{n}$, $\bar{a}_{n}$, $b_{n}$, $\bar{b}_{n}$, $f_{n}$, $g_{n}$ ($n\in\mathbf{Z_{>0}}$), $x$, are Grassmann-even numbers. \\[2mm]
{\tt [Formula\hspace{0.7mm}1]}
\be
{}_{f_{n}} \langle a_{n},\bar a_{n}|b_{n},\bar b_{n}\rangle_{g_{n}}
&:=&\langle 0|\exp\!\left[\frac{f_{n}}{n}\alpha_{n}\widetilde{\alpha}_{n}\right]\exp[\alpha_{n}a_{n}+\widetilde{\alpha}_{n}\bar{a}_{n}]\exp[b_{n}\alpha_{-n}+\bar{b}_{n}\widetilde{\alpha}_{-n}]\exp\!\left[\frac{g_{n}}{n}\alpha_{-n}\widetilde{\alpha}_{-n}\right]|0\rangle \\[3mm]
&=&\frac{1}{1-f_{n}g_{n}}\hspace{1mm}\exp\!\bigg[\frac{n}{1-f_{n}g_{n}}(a_{n}b_{n}+\bar{a}_{n}\bar{b}_{n}+g_{n}a_{n}\bar{a}_{n}+f_{n}b_{n}\bar{b}_{n})\bigg] \\
\ee
{\tt ex.)}
$$
\langle T_{v}(K)|e^{-s(L_{0}+\widetilde{L}_{0})}|T_{u}(K)\rangle
=\prod_{n=1}^{\infty} {}_{f_{n}} \langle a_{n},\bar{a}_{n}|b_{n},\bar{b}_{n}\rangle_{g_{n}}
$$ \\[5mm]
{\tt [Formula\hspace{0.7mm}2]}
\be
\frac{{}_{f_{n}} \langle a_{n},\bar{a}_{n}|\alpha_{n}|b_{n},\bar{b}_{n}\rangle_{g_{n}}}{{}_{f_{n}} \langle a_{n},\bar{a}_{n}|b_{n},\bar{b}_{n}\rangle_{g_{n}}}
&=&{{d\over dx} \hspace{1mm} {}_{f_{n}}\langle a_{n}+x,\bar a_{n}|b_{n},\bar b_{n}\rangle_{g_{n}}|_{x=0} \over {}_{f_{n}} \langle a_{n},\bar a_{n}|b_{n},\bar b_{n}\rangle_{g_{n}}}
={n(b_{n}+g_{n}\bar a_{n}) \over 1-f_{n}g_{n}} \\[5mm]
\frac{{}_{f_{n}} \langle a_{n},\bar{a}_{n}|\widetilde{\alpha}_{n}|b_{n},\bar{b}_{n}\rangle_{g_{n}}}{{}_{f_{n}} \langle a_{n},\bar{a}_{n}|b_{n},\bar{b}_{n}\rangle_{g_{n}}}
&=&{{d\over dx} \hspace{1mm} {}_{f_{n}}\langle a_{n},\bar a_{n}+x|b_{n},\bar b_{n}\rangle_{g_{n}}|_{x=0} \over {}_{f_{n}} \langle a_{n},\bar a_{n}|b_{n},\bar b_{n}\rangle_{g_{n}}}
={n(\bar b_{n}+g_{n}a_{n}) \over 1-f_{n}g_{n}} \\[5mm]
\frac{{}_{f_{n}} \langle a_{n},\bar{a}_{n}|\alpha_{-n}|b_{n},\bar{b}_{n}\rangle_{g_{n}}}{{}_{f_{n}} \langle a_{n},\bar{a}_{n}|b_{n},\bar{b}_{n}\rangle_{g_{n}}}
&=&{{d\over dx} \hspace{1mm} {}_{f_{n}}\langle a_{n},\bar a_{n}|b_{n}+x,\bar b_{n}\rangle_{g_{n}}|_{x=0} \over {}_{f_{n}} \langle a_{n},\bar a_{n}|b_{n},\bar b_{n}\rangle_{g_{n}}}
={n(a_{n}+f_{n}\bar b_{n}) \over 1-f_{n}g_{n}} \\[5mm]
\frac{{}_{f_{n}} \langle a_{n},\bar{a}_{n}|\widetilde{\alpha}_{-n}|b_{n},\bar{b}_{n}\rangle_{g_{n}}}{{}_{f_{n}} \langle a_{n},\bar{a}_{n}|b_{n},\bar{b}_{n}\rangle_{g_{n}}}
&=&{{d\over dx} \hspace{1mm} {}_{f_{n}}\langle a_{n},\bar a_{n}|b_{n},\bar b_{n}+x\rangle_{g_{n}}|_{x=0} \over {}_{f_{n}} \langle a_{n},\bar a_{n}|b_{n},\bar b_{n}\rangle_{g_{n}}}
={n(\bar a_{n}+f_{n}b_{n}) \over 1-f_{n}g_{n}} \\
\ee
{\tt ex.)}
$$
\langle \alpha_{n}\rangle
\equiv\lim_{s \to 0}\frac{\langle T_{v}(K)|\alpha_{n}\hspace{0.7mm}e^{-s(L_{0}+\widetilde{L}_{0})}|T_{u}(K)\rangle}{\langle T_{v}(K)|e^{-s(L_{0}+\widetilde{L}_{0})}|T_{u}(K)\rangle}\bigg|_{u=v} 
=\lim_{s \to 0}\frac{{}_{f_{n}} \langle a_{n},\bar{a}_{n}|\alpha_{n}|b_{n},\bar{b}_{n}\rangle_{g_{n}}}{{}_{f_{n}} \langle a_{n},\bar{a}_{n}|b_{n},\bar{b}_{n}\rangle_{g_{n}}}
=\lim_{s \to 0}\frac{n(b_{n}+g_{n}\bar{a}_{n})}{1-f_{n}g_{n}}
$$ \\[1cm]
{\bf [Fermionic part]} \\[2mm]
$a_{r}$, $\bar{a}_{r}$, $b_{r}$, $\bar{b}_{r}$ ($r\in\mathbf{Z}_{\geq 0}+\frac{1}{2}$), $\theta$ are Grassmann-odd and $f_{r}$, $g_{r}$ ($r\in\mathbf{Z}_{\geq 0}+\frac{1}{2}$) are Grassmann-even numbers. \\[2mm]
{\tt [Formula\hspace{0.7mm}1]}
\begin{eqnarray*}
{}_{f_{r}} \langle a_{r},\bar a_{r}|b_{r},\bar b_{r}\rangle_{g_{r}}
&:=&\langle 0|\exp[f_{r}\psi_{r}\widetilde{\psi}_{r}]\exp[\psi_{r}a_{r}+\widetilde{\psi}_{r}\bar{a}_{r}]\exp[b_{r}\psi_{-r}+\bar{b}_{r}\widetilde{\psi}_{-r}]\exp[g_{r}\psi_{-r}\widetilde{\psi}_{-r}]|0\rangle \\[3mm]
&=&(1-f_{r}g_{r})\hspace{1mm}\exp\!\bigg[\frac{1}{1-f_{r}g_{r}}(a_{r}b_{r}+\bar{a}_{r}\bar{b}_{r}+g_{r}a_{r}\bar{a}_{r}+f_{r}b_{r}\bar{b}_{r})\bigg] \\[3mm]
\end{eqnarray*}
{\tt [Formula\hspace{0.7mm}2]}
\begin{eqnarray*}
\frac{{}_{f_{r}} \langle a_{r},\bar{a}_{r}|\psi_{r}|b_{r},\bar{b}_{r}\rangle_{g_{r}}}{{}_{f_{r}} \langle a_{r},\bar{a}_{r}|b_{r},\bar{b}_{r}\rangle_{g_{r}}}
&=&{{d\over d\theta} \hspace{1mm} {}_{f_{r}}\langle a_{r}+\theta,\bar a_{r}|b_{r},\bar b_{r}\rangle_{g_{r}} \over {}_{f_{r}} \langle a_{r},\bar a_{r}|b_{r},\bar b_{r}\rangle_{g_{r}}}
={-b_{r}-g_{r}\bar a_{r} \over 1-f_{r}g_{r}} \\[5mm]
\frac{{}_{f_{r}} \langle a_{r},\bar{a}_{r}|\widetilde{\psi}_{r}|b_{r},\bar{b}_{r}\rangle_{g_{r}}}{{}_{f_{r}} \langle a_{r},\bar{a}_{r}|b_{r},\bar{b}_{r}\rangle_{g_{r}}}
&=&{{d\over d\theta} \hspace{1mm} {}_{f_{r}} \langle a_{r},\bar a_{r}+\theta|b_{r},\bar b_{r}\rangle_{g_{r}} \over {}_{f_{r}} \langle a_{r},\bar a_{r}|b_{r},\bar b_{r}\rangle_{g_{r}}}
={-\bar b_{r}+g_{r}a_{r} \over 1-f_{r}g_{r}} \\[5mm]
\frac{{}_{f_{r}} \langle a_{r},\bar{a}_{r}|\psi_{-r}|b_{r},\bar{b}_{r}\rangle_{g_{r}}}{{}_{f_{r}} \langle a_{r},\bar{a}_{r}|b_{r},\bar{b}_{r}\rangle_{g_{r}}}
&=&{{d\over d\theta}\hspace{1mm} {}_{f_{r}} \langle a_{r},\bar a_{r}|b_{r}+\theta,\bar b_{r}\rangle_{g_{r}} \over {}_{f_{r}} \langle a_{r},\bar a_{r}|b_{r},\bar b_{r}\rangle_{g_{r}}}
={-a_{r}+f_{r}\bar b_{r} \over 1-f_{r}g_{r}} \\[5mm]
\frac{{}_{f_{r}} \langle a_{r},\bar{a}_{r}|\widetilde{\psi}_{-r}|b_{r},\bar{b}_{r}\rangle_{g_{r}}}{{}_{f_{r}} \langle a_{r},\bar{a}_{r}|b_{r},\bar{b}_{r}\rangle_{g_{r}}}
&=&{{d\over d\theta}\hspace{1mm} {}_{f_{r}} \langle a_{r},\bar a_{r}|b_{r},\bar b_{r}+\theta \rangle_{g_{r}} \over {}_{f_{r}} \langle a_{r},\bar a_{r}|b_{r},\bar b_{r}\rangle_{g_{r}}}
={-\bar a_{r}-f_{r}b_{r} \over 1-f_{r}g_{r}} \\[2mm]
\end{eqnarray*}
{\bf Appendix B}

We show a proper regularization of the infinite product $\langle \Phi \rangle := \prod_{n=1}^{\infty}\prod_{r=\frac{1}{2}}^{\infty}\left(\frac{n}{r}\right)$ (\,eq.(\ref{vev of tachyon field})\,) as $\sqrt{\pi}$.
This is the vev of the tachyon field $\Phi_{u,\rho}$ in the case of $(u,\rho)=(\infty,+1)$. 

The inverse of the infinite product $\langle\Phi\rangle^{-1}$ is written in two ways as
\begin{eqnarray*}
\langle\Phi\rangle^{-1} 
&=&\prod_{n=1}^\infty{n-{1\over 2}\over n}=\prod_{n=1}^\infty\left(1-{1\over 2n}\right) \\[2mm]
&=&{1\over 2}\prod_{n=1}^\infty{n+{1\over 2}\over n}={1\over 2}\prod_{n=1}^\infty\left(1+{1\over 2n}\right).
\end{eqnarray*}
We take the product of the former and the latter and calculate it as
$$ \langle\Phi\rangle^{-2}={1\over 2}\prod_{n=1}^\infty\left(1-{1\over 2n}\right)\left(1+{1\over 2n}\right)
={1\over \pi}, $$
where we used the following formula
$$ \prod_{n=1}^\infty\left(1-{x\over n}\right)\left(1+{x\over n}\right)={\sin \pi x\over \pi x}. $$ 
Thus $\langle\Phi\rangle$ is properly regularized as $$\langle\Phi\rangle = \sqrt{\pi},$$
Note that the value depends on the order of the products.
\end{document}